\documentclass[12pt,english,floatfix,nofootinbib,superscriptaddress,aps,prd,preprint]{revtex4}
\usepackage[utf8]{inputenc}       
\usepackage[english]{babel}       
\usepackage[utf8]{inputenc}
\usepackage{textcomp}
\usepackage{amsmath,amsthm,amssymb,amsfonts,mathrsfs,amsbsy} 
\usepackage{tensor}               
\usepackage{slashed}  
\usepackage{bbm}
\usepackage{esint}                
\usepackage{cancel}               
\usepackage{tikz}
\usetikzlibrary{arrows.meta,positioning,calc}
\usepackage{graphicx}             
\usepackage{natbib}
\usepackage{float}                
\usepackage{subfig}               
\usepackage[font=small,labelfont=bf]{caption} 
\usepackage{multirow}             
\usepackage{array}                
\usepackage{tikz}                 
\usepackage{subcaption}
\usetikzlibrary{quotes,angles,arrows,decorations.markings} 
\usepackage{makecell} 

\usepackage{lipsum}

\usepackage{xcolor}               
\usepackage{color}                
\usepackage{textcomp}             
\usepackage{units}                

\newcommand{\be}{\begin{align}}
\newcommand{\ee}{\end{align}}
\newcommand{\Be}{\begin{eqnarray}}
\newcommand{\Ee}{\end{eqnarray}}

\newcommand{\mincir}{\raise
-3.truept\hbox{\rlap{\hbox{$\sim$}}\raise4.truept\hbox{$<$}\ }}
\newcommand{\magcir}{\raise
-3.truept\hbox{\rlap{\hbox{$\sim$}}\raise4.truept\hbox{$>$}\ }}

\newcolumntype{Y}{>{\centering\arraybackslash}X}
\providecommand{\U}[1]

\usepackage[dvips]{epsfig}        
\usepackage[dvips]{graphicx}      
\usepackage{hyperref}             
\hypersetup{
    colorlinks=true,              
    breaklinks=true,              
    citecolor=blue,               
    linkcolor=[rgb]{0,0.5,0.9},   
    urlcolor=red,                 
    filecolor=green               
}

\newcommand{\ie}{\begin{align}}
\newcommand{\fe}{\end{align}}
\newcommand{\se}{\begin{eqnarray}}
\newcommand{\ff}{\end{eqnarray}}

\begin{document}

\title{Accretion of matter of a new bumblebee black hole}


\author{Yuxuan Shi}
\email{shiyx2280771974@gmail.com}
\affiliation{Department of Physics, East China University of Science and Technology, Shanghai 200237, China}


\author{A. A. Ara\'{u}jo Filho}
\email{dilto@fisica.ufc.br}
\affiliation{Departamento de Física, Universidade Federal da Paraíba, Caixa Postal 5008, 58051--970, João Pessoa, Paraíba,  Brazil.}
\affiliation{Departamento de Física, Universidade Federal de Campina Grande Caixa Postal 10071, 58429-900 Campina Grande, Paraíba, Brazil.}
\affiliation{Center for Theoretical Physics, Khazar University, 41 Mehseti Street, Baku, AZ-1096, Azerbaijan.}



\date{\today}

\begin{abstract}

We investigate how the newly obtained static black hole in bumblebee gravity affects the behavior of accreting matter and its observable signatures. The Lorentz--violating parameter that characterizes this geometry modifies photon trajectories and shifts the location of the critical curve that defines the shadow. Using ray tracing, we examine light deflection, the structure of direct emission, lensing rings, and photon rings, and we explore three thin--disk emission models—starting at the ISCO, at the photon sphere, and at the event horizon—together with static and infalling spherical accretions. Larger values of this parameter enlarge the shadow, move all optical features outward, and suppress the observed intensity through gravitational redshift, with additional dimming produced by Doppler effects for infalling matter.

\end{abstract}


\maketitle

\tableofcontents


\section{Introduction}

Ideas suggesting that Lorentz symmetry may operate only as an approximate feature of spacetime rather than an exact rule have become increasingly relevant, especially in scenarios that aim to connect gravitation with quantum physics \cite{colladay1997cpt,kostelecky1989spontaneous,kostelecky2004gravity,kostelecky2011data,kostelecky1999constraints}. Many of these approaches anticipate that new geometric effects could emerge near experimentally reachable energy scales. In such contexts, departures from perfect Lorentz invariance often appear when particular dynamical fields naturally settle into vacuum states with nonzero values, establishing a preferred spacetime direction and triggering spontaneous symmetry breaking. Bumblebee constructions offer a concise way to implement this idea. These models introduce a vector field whose norm is fixed through a potential, forcing the field to stabilize at a constant magnitude. Once the vacuum solution is reached, this background vector endows the spacetime with an orientation that alters the standard relativistic structure. The modified geometry behaves consistently within the theoretical framework and supplies a well--organized description of Lorentz--violating effects \cite{bluhm2005spontaneous,Bluhm:2023kph,Bluhm:2019ato,Maluf:2014dpa,Maluf:2013nva,bluhm2008spontaneous}.

A variety of theoretical programs—ranging from extensions of Einstein’s gravity to string-motivated scenarios—have highlighted situations in which a background vector field may alter the local spacetime symmetries \cite{kostelecky1989spontaneous,kostelecky1991photon,jacobson2004einstein}. Within these approaches, effective descriptions often contain fields that naturally stabilize at nonvanishing vacuum values, and once this occurs, the relativistic symmetry of the geometry no longer remains exact \cite{kostelecky2004gravity,bluhm2005spontaneous}. Bumblebee models offer one of a remarkable approach concerning this idea. In such constructions, the central ingredient is a vector quantity $B_{\mu}$ whose magnitude is regulated by a potential of the form $V(B_{\mu} B^{\mu} \mp b^{2})$ \cite{Liu:2022dcn}. The potential constrains the field to approach a configuration with fixed norm, and when this condition is met, the resulting vacuum establishes a preferred direction in spacetime. This selection mechanism realizes the spontaneous breaking of Lorentz symmetry \cite{bluhm2005spontaneous,bluhm2008spontaneous}. Once the vacuum background is present, perturbations split into two qualitatively distinct sectors. Disturbances that preserve the fixed--norm condition propagate similarly to massless gauge modes and resemble photon--like excitations \cite{bluhm2005spontaneous}, while those that modify the constrained magnitude obtain mass through the same potential that enforces the vacuum configuration \cite{bluhm2008spontaneous}.

Once the bumblebee mechanism was incorporated into curved geometries, the vacuum configuration of the vector field necessarily interacted with the spacetime dynamics, giving rise to a broad set of gravitational applications \cite{Bertolami:2005bh}. This integration opened multiple research paths that evolved in parallel. One prominent direction focused on compact objects. The black hole metric introduced in \cite{Casana:2017jkc} became the baseline for examining gravitational phenomena in backgrounds where Lorentz symmetry is broken. Studies employing this solution analyzed how near--horizon physics is altered, including changes in entanglement behavior \cite{Liu:2024wpa} and the structure of particle production in these deformed geometries \cite{AraujoFilho:2025hkm}. A related line of work adapted the mechanism to tensorial fields within the Kalb--Ramond setting, producing a broader class of Lorentz-violating black hole configurations \cite{AraujoFilho:2024ctw}. A different set of developments moved toward cosmology and astrophysical modeling. Scenarios that mimic anisotropic expansion patterns similar to Kasner--type universes were constructed in \cite{Neves:2022qyb}, while \cite{Neves:2024ggn} examined how the same vector background alters the internal structure of anisotropic stellar systems. The propagation of gravitational waves was also revisited in this context, showing departures from predictions of standard general relativity \cite{Liang:2022hxd,amarilo2024gravitational}.
Additional generalizations modified the geometric sector more directly, for instance by incorporating a cosmological constant into the bumblebee framework and analyzing the resulting relaxed vacuum structures \cite{Maluf:2020kgf}.

Over the past several years, the bumblebee framework has expanded far beyond the initial static background of Ref.~\cite{Casana:2017jkc}. One of the most active developments emerged within the \textit{metric--affine} approach, where the independent connection introduces new geometric degrees of freedom. Within this setting, a static configuration was derived in \cite{Filho:2022yrk}, followed by the construction of a rotating counterpart displaying axial symmetry \cite{AraujoFilho:2024ykw}. These advances later enabled the formulation of non--commutative versions of the theory \cite{AraujoFilho:2025rvn} together with analogous extensions in tensor--field frameworks such as Kalb–Ramond gravity \cite{AraujoFilho:2025jcu}.  Furthermore, attention has not remained restricted to black holes. A growing line of research has shown that the fixed--norm vector field can sustain wormhole geometries or influence their traversability properties \cite{Ovgun:2018xys,Magalhaes:2025nql,Magalhaes:2025lti,AraujoFilho:2024iox}. Parallel investigations have proposed black--bounce backgrounds driven by $\kappa$–essence fields while still incorporating symmetry--breaking effects \cite{Pereira:2025xnw}. Propagation phenomena provide another active domain. The deflection of neutrinos has been examined across several realizations of the theory: purely metric versions \cite{Shi:2025plr}, formulations with independent metric and connection \cite{Shi:2025ywa}, and tensor--based generalizations of the bumblebee sector \cite{Shi:2025rfq}.

Recent developments have expanded the studies of Lorentz--violating black holes by introducing new solutions derived directly from the bumblebee mechanism \cite{Liu:2025oho,Zhu:2025fiy}. A detailed analysis of this new static member of this family was later carried out in \cite{AraujoFilho:2025zaj}, where several gravitational aspects and constrainsts were investigated. More recently, the implications of this background for neutrino propagation—specifically oscillation phenomena—were analyzed in \cite{Shi:2025tvu}. A rotating analogue of the solution has also become available. By applying an improved Newman--Janis transformation to the static seed, an axisymmetric geometry was obtained and presented in \cite{Kumar:2025bim}.

When photons pass near a compact object, their trajectories are warped so strongly that a characteristic dark patch appears against any illuminating background. This absence of light—identified today as the black hole shadow—arises because part of the incoming radiation becomes trapped on unstable photon orbits while the remainder is bent away so severely that it fails to reach the observer. The notion that such a silhouette could reveal properties of strong gravitational fields dates back to the seminal analyses of the 1970s, where Bardeen and others examined how rotating black hole structures ~\cite{Cunningham}. These early theoretical features later matured into concrete observational strategies. Falcke, Melia, and Agol argued that the compact object at the center of the Galaxy should produce a detectable depression in the submillimeter emission~\cite{Falcke:1999pj}. Their proposal eventually guided the long--term development of horizon--scale interferometry, culminating in the Event Horizon Telescope program. This effort ultimately delivered the first resolved image of a black hole in M87, followed by the reconstruction of the shadow of $Sgr\,A^{*}$. The availability of such data transformed studies of strong gravity. Predictions of shadow sizes, deformations, and asymmetries—within general relativity and numerous alternative theories—are now routinely brought into comparison with observational bounds. A rapidly growing literature examines how different gravitational models modify the predicted silhouette, using shadow observables as discriminating tools~\cite{Afrin:2024khy,Khodadi:2024ubi,Allahyari:2019jqz,Afrin:2021wlj,Nojiri:2024txy,Afrin:2021imp,Nojiri:2024qgx,Bambi:2019tjh,Khodadi:2021gbc,Liu:2024lve,Khodadi:2022pqh,Kumar:2020hgm,Vagnozzi:2019apd,Nojiri:2024nlx,Fu:2021fxn,Liu:2024soc}.

A meaningful way to interpret images of compact objects—and to look for departures from standard gravitational behavior—is to combine the spacetime geometry with a simplified description of the surrounding accretion flow. Earlier analyses considered Schwarzschild black holes illuminated by spherically symmetric infall or by alternative accretion prescriptions. Those studies revealed a striking feature: the dark region associated with the shadow remains remarkably stable, its shape and radius governed almost entirely by the underlying metric rather than by the detailed properties of the accreting material \cite{40,41,42}. When more realistic configurations were explored, including optically thin or geometrically thin disks, a different aspect became evident. In such scenarios, the apparent size of the shadow is influenced by the location and distribution of the emitting matter \cite{Gogoi:2024eyw,Zare:2024dtf,43,44,46,47,48,49,AraujoFilho:2024mvz,Lambiase:2023zeo}. The bright domain surrounding the dark core is typically composed of several visual structures: direct radiation from the disk, highly bent photons that form lensing arcs, and the increasingly narrow photon rings produced by repeated near--critical orbits \cite{Macedo:2025ipc,Olmo:2025ctf,Rosa:2024eva,Macedo:2024qky}.

Although previous works examined several classical and semiclassical properties of this solution, its impact on accreting matter has not yet been addressed. The present analysis fills this gap by examining how the recently derived static black hole in bumblebee gravity alters the dynamics of accretion and the resulting observational appearance. We analyze the influence of the Lorentz--violating parameter on photon trajectories and on the position of the critical curve that defines the shadow. By means of ray tracing, we study light bending, the distribution of direct emission, and the structure of lensing and photon rings. Three thin--disk configurations are considered—starting at the ISCO, at the photon sphere, and at the event horizon—together with static and radially infalling spherical accretions. Increasing values of this parameter expand the shadow, shift all optical structures outward, and reduce the observed brightness through gravitational redshift, with additional suppression arising from Doppler effects in the infalling cases.


\section{The new bumblebee solution}

A recently introduced classes of Lorentz--violating compact objects has been modeled through a modified geometry obtained within the bumblebee framework. The metric that characterizes this new black hole configuration was presented in Refs.~\cite{Liu:2025oho,Zhu:2025fiy}
\begin{align}
\label{metric}
\mathrm{d}s^{2} = - \dfrac{1}{1+\chi}\left(1 - \dfrac{2M}{r}\right)\mathrm{d}t^{2} + (1+\chi)\left(1 - \dfrac{2M}{r} \right)^{-1} \mathrm{d}r^{2} + r^{2}\mathrm{d}\Omega^{2}.
\end{align}
In this geometry, the deviation from standard Lorentz symmetry is encoded in the parameter $\chi$, which combines two distinct ingredients of the model through the relation $\chi = \alpha\,\ell$. The quantity $\alpha$ arises as an integration constant of the solution, while $\ell$ is determined by the background configuration of the bumblebee field. More precisely, $\ell$ is defined as $\ell = \xi\, b^{2}$, where $b^{2} = b_{\mu} b^{\mu}$ denotes the fixed norm selected by the vacuum of the vector field and $\xi$ represents the nonminimal coupling that links the vector background to the curvature.

A natural question arises at this stage: if we rescale the temporal coordinate so as to absorb the constant factor in $g_{tt}$, would the resulting geometry become equivalent to the usual bumblebee black hole \cite{Casana:2017jkc}, in which only $g_{rr}$ is modified? The answer is no. In the bumblebee construction, the metric and the vacuum configuration $b_{\mu}$ are obtained simultaneously from the field equations under the constraint $b_{\mu}b^{\mu}=s\,b^{2}$, and the constant prefactor in $g_{tt}$ is fixed by the same parameters that determine the functional form of $b_{t}(r)$ and $b_{r}(r)$. This factor is therefore not a removable normalization, but a physical element tied to the Lorentz--violating background. If one “absorbs’’ this constant factor into $g_{tt}$ while keeping $b_{\mu}$ fixed in the coordinate basis (as required when deriving the black hole solution), the VEV constraint is no longer satisfied because $b_{\mu}$ will not being transformed as a covector.

Let us emphasize that this only happens because we are not performing a genuine coordinate transformation. If one actually changes coordinates from $t$ to $\tau$ by
\begin{equation}
t = \sqrt{K}\,\tau,
\end{equation}
then both the metric and the components of $b_{\mu}$ must transform. Since $b_{\mu}$ is a covector, one straightforwardly obtains
\begin{equation}
\hat{b}_{\tau}
= \frac{\partial t}{\partial \tau}\,b_{t}
= \sqrt{K}\,b_{t},
\qquad
\hat{b}_{r} = b_{r},
\end{equation}
while the inverse metric components in the new coordinates become
\begin{equation}
\hat{g}^{\tau\tau} = -\frac{1}{f(r)},
\qquad
\hat{g}^{rr} = \frac{f(r)}{K}.
\end{equation}
The norm of the bumblebee field then reads
\begin{equation}
\hat{b}_{\mu}\hat{b}^{\mu}
= \hat{g}^{\tau\tau}\hat{b}_{\tau}^{2}
+ \hat{g}^{rr}\hat{b}_{r}^{2}
= -\frac{1}{f}\,\bigl(\sqrt{K}\,b_{t}\bigr)^{2}
+ \frac{f}{K}\,b_{r}^{2}
= -\frac{K}{f}\,b_{t}^{2}
+ \frac{f}{K}\,b_{r}^{2}
= b_{\mu}b^{\mu}.
\end{equation}
In oher words, under a true coordinate change in which both $g_{\mu\nu}$ and $b_{\mu}$ transform as tensorial objects, the quantity $b_{\mu}b^{\mu}$ is invariant, exactly as expected.

However, when one aims at obtaining the black hole solutions of the bumblebee theory, the functional form of $b_{\mu}$ must remain fixed, because it is determined simultaneously with the metric by the field equations and the VEV constraint. Therefore, modifying only the metric while keeping $b_{\mu}$ unchanged in the coordinate basis leads to inconsistencies in the basic structure of the solution. To see this explicitly, consider the original line element
\begin{equation}
\mathrm{d}s^{2}
= -\frac{1}{K}\,f(r)\,\mathrm{d}t^{2}
+ K\,f(r)^{-1}\,\mathrm{d}r^{2}
+ r^{2}\mathrm{d}\Omega^{2},
\qquad K = 1+\chi,
\end{equation}
whose inverse metric components are
\begin{equation}
g^{tt} = -\frac{K}{f(r)}, 
\qquad
g^{rr} = \frac{f(r)}{K}.
\end{equation}
With $b_{\mu} = \bigl(b_{t}(r),b_{r}(r),0,0\bigr)$, the VEV condition reads
\begin{equation}
b_{\mu}b^{\mu}
= g^{tt}b_{t}^{2}+g^{rr}b_{r}^{2}
= -\frac{K}{f}\,b_{t}^{2}+\frac{f}{K}\,b_{r}^{2}
= s\,b^{2}.
\tag{1}
\end{equation}

Now perform only the temporal rescaling
\begin{equation}
t = \sqrt{K}\,\tau,
\qquad
\mathrm{d}t^{2} = K\,\mathrm{d}\tau^{2},
\end{equation}
leaving $r$ unchanged. The metric becomes
\begin{equation}
\widehat{\mathrm{d}s}^{2}
= - f(r)\,\mathrm{d}\tau^{2}
+ K\,f(r)^{-1}\,\mathrm{d}r^{2}
+ r^{2}\mathrm{d}\Omega^{2}.
\end{equation}
If we now keep the same functions $b_{t}(r)$ and $b_{r}(r)$ (i.e.\ we do not transform $b_{\mu}$ as a covector), the new norm becomes
\begin{equation}
\hat{b}_{\mu}\hat{b}^{\mu}
= -\frac{1}{f}\,b_{t}^{2}
+ \frac{f}{K}\,b_{r}^{2}.
\end{equation}
Then
\begin{equation}
\hat{b}_{\mu}\hat{b}^{\mu} - b_{\mu}b^{\mu}
= \frac{K-1}{f(r)}\,b_{t}^{2}.
\end{equation}
For $b_{t}\neq 0$ and $K\neq 1$ (i.e.\ $\chi\neq0$), this correction does not vanish, and therefore
\begin{equation}
\hat{b}_{\mu}\hat{b}^{\mu}\neq s\,b^{2}.
\end{equation}

\begin{figure}[t]
\centering
\begin{tikzpicture}[
    node distance=1.3cm and 2.0cm,
    >=Latex,
    box/.style={
      draw,
      rounded corners,
      align=center,
      inner sep=3pt,
      font=\small,
      minimum width=4.2cm
    }
  ]

  \node[box] (orig) {Original BH solution\\[2pt]
    $(g_{\mu\nu}, b_\mu)$\\[2pt]
    $b_\mu b^\mu = s\,b^2$};

  \node[box, below left=0.9cm and 1.6cm of orig] (left1) {Genuine coordinate transformation\\[2pt]
    $t \rightarrow \sqrt{K}\,\tau$\\
    $g_{\mu\nu}\rightarrow \hat{g}_{\mu\nu}$\\
    $b_\mu \rightarrow \hat{b}_\mu$};
  \node[box, below=0.8cm of left1] (left2) {Norm preserved:\\[2pt]
    $\hat{b}_\mu \hat{b}^\mu = b_\mu b^\mu$\\
    Same physical configuration};

  \node[box, below right=0.9cm and 1.6cm of orig] (right1) {Metric-only rescaling\\[2pt]
    $t \rightarrow \sqrt{K}\,\tau$\\
    $g_{\mu\nu}\rightarrow \hat{g}_{\mu\nu}$\\
    $b_\mu$ kept fixed};
  \node[box, below=0.8cm of right1] (right2) {VEV constraint broken:\\[2pt]
    $\hat{b}_\mu \hat{b}^\mu \neq s\,b^2$\\
    Not a valid solution};

  \coordinate (midLR) at ($(left2.south)!0.5!(right2.south)$);

  \node[box, below=0.8cm of midLR] (bottom) {\underline{Bumblebee BH construction}:\\[2pt]
    $b_\mu(r)$  {\bf{must be fixed}} by the VEV ansatz\\
    and the field equations.\\[2pt]
    Keeping $b_\mu$ fixed while changing only $g_{\mu\nu}$\\
    breaks $b_\mu b^\mu = s\,b^2$.\\[2pt]
    Transforming $b_\mu$ {\bf{covariantly}} preserves the norm,\\
    but {\bf{changes the vacuum configuration}},\\
    corresponding to a {\bf{different BH}} solution};

  \draw[->] (orig.south west) -- (left1.north);
  \draw[->] (orig.south east) -- (right1.north);
  \draw[->] (left1.south) -- (left2.north);
  \draw[->] (right1.south) -- (right2.north);
  \draw[->] (left2.south) -- (bottom.north west);
  \draw[->] (right2.south) -- (bottom.north east);

\end{tikzpicture}

\caption{Schematic comparison between two distinct operations starting from the original bumblebee black hole solution. Left: a genuine coordinate transformation rescales $t$ and transforms both the metric and the covector $b_\mu$, preserving the invariant norm $b_\mu b^\mu$ and describing the same physical configuration in a different coordinate chart. Right: absorbing the prefactor in $g_{tt}$ while keeping $b_\mu$ fixed in the coordinate basis breaks the VEV constraint, producing a configuration that does not satisfy the bumblebee field equations. The bottom box summarizes the tension: in the construction of the bumblebee black hole, $b_\mu$ is fixed by the VEV ansatz and the equations of motion; changing only the metric is inconsistent, while transforming $b_\mu$ covariantly preserves covariance but corresponds to a different vacuum configuration and therefore to a different black hole solution.}
\label{fig:bumblebee-transformation}
\end{figure}
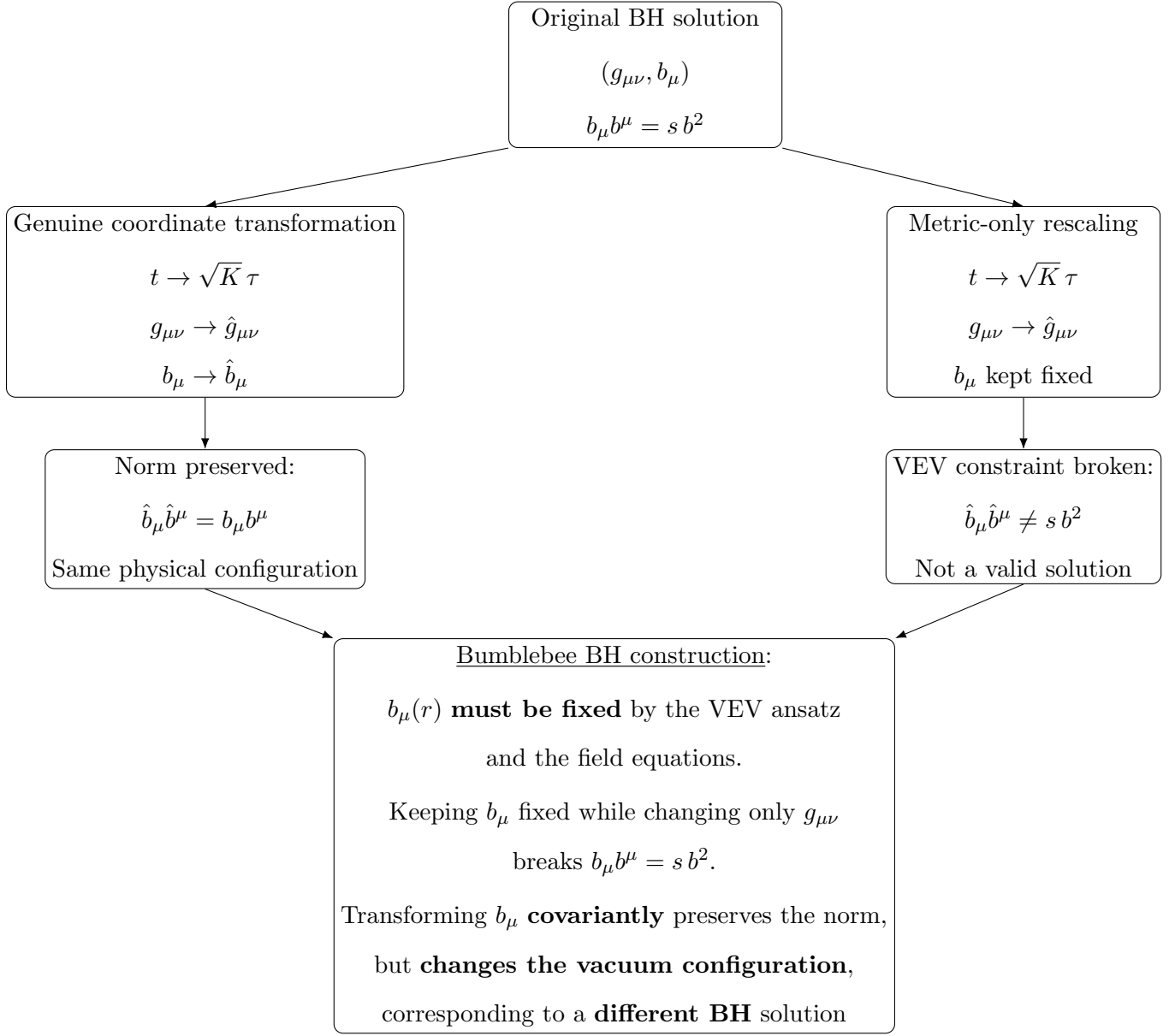

In other words, absorbing the constant prefactor in $g_{tt}$ while keeping $b_{\mu}$ fixed in the coordinate basis (as required when deriving the black hole solution) breaks the VEV constraint and does not reproduce the original black hole configuration. A genuine coordinate transformation would preserve $b_{\mu}b^{\mu}$ by transforming both $g_{\mu\nu}$ and $b_{\mu}$, but this necessarily modifies the functional form of $b_{\mu}$, so the resulting configuration no longer corresponds to the same vacuum determined by the field equations and therefore does not describe the same black hole. The points discussed above are concisely illustrated in the schematic diagram shown in Fig.~\ref{fig:bumblebee-transformation}.

\section{Light deflection in the new bumblebee black hole}

We examine photon mobility in this spacetime to explore the optical appearance of the black hole, namely its shadow and the corresponding pictures of the accretion disc. The null geodesics, which may be obtained from the Euler--Lagrange equations, control how light propagates:
\begin{align}
\dfrac{\mathrm{d}}{\mathrm{d}\lambda}\left(\dfrac{\partial\mathcal{L}}{\partial\dot{x}^{\mu}}\right)=\dfrac{\partial\mathcal{L}}{\partial x^{\mu}},
\end{align}
where $\lambda$ denotes the affine parameter and $\dot{x}^{\mu}$ represents the four--velocity of the photon. The Lagrangian $\mathcal{L}$ for a massless particle simplifies when we limit our analysis to the equatorial plane ($\theta = \pi/2$), and it is clearly written as \cite{synge1966escape,bardeen1972rotating,gralla2020lensing}
\begin{align}
2\mathcal{L} = - \dfrac{1}{1+\chi}\left(1 - \dfrac{2M}{r}\right)\dot{t}^{2} + (1+\chi)\left(1 - \dfrac{2M}{r}\right)^{-1}\dot{r}^{2} + r^{2}\dot{\phi}^{2} = 0,
\end{align}
where the dot denotes the derivative with respect to $\lambda$. The spacetime has two Killing vectors, $\partial_t$ and $\partial_\phi$, due to stationarity and axial symmetry. Two conserved values result from these symmetries: the photon's energy $E$ and orbital angular momentum $L$, which are determined by \cite{shaikh2019shadows,he2022shadow,li2021shadows,shi2024shadow,feng2025shadow}:
\begin{align}
E &= -\dfrac{\partial \mathcal{L}}{\partial \dot{t}} = \dfrac{1}{1+\chi}\left(1 - \dfrac{2M}{r}\right)\dot{t}, \\
L &= \dfrac{\partial \mathcal{L}}{\partial \dot{\phi}} = r^{2}\dot{\phi}.
\end{align}
To facilitate the derivation, we introduce the impact parameter $b \equiv |L|/E$ and rescale the affine parameter via $\tau = \lambda/|L|$. Consequently, the equations for the time and azimuthal components can be rewritten as \cite{amir2018shadows,eiroa2018shadow,shaikh2019shadows,he2022shadow,li2021shadows}
\begin{align}
\dfrac{\mathrm{d}t}{\mathrm{d}\tau} &= \dfrac{1+\chi}{b}\left(1-\dfrac{2M}{r}\right)^{-1}, \\
\dfrac{\mathrm{d}\phi}{\mathrm{d}\tau} &= \pm \dfrac{1}{r^{2}},
\end{align}
where the sign $\pm$ determines the direction of the light ray (counter--clockwise or clockwise). Substituting these expressions back into the null constraint $2\mathcal{L} = 0$, we obtain the radial equation of motion:
\begin{align}
\left(\dfrac{\mathrm{d}r}{\mathrm{d}\tau}\right)^{2} + V_{\mathrm{eff}}(r) = \dfrac{1}{b^2}.
\end{align}
Here, $V_{\mathrm{eff}}(r)$ is the effective potential modified by the bumblebee parameter $\chi$. A consistent derivation from the metric Eq. \eqref{metric} yields the explicit form \cite{gralla2020lensing,li2021observational,fathi2023observational,he2022shadow,shi2024shadow}:
\begin{align}
\label{Veff}
V_{\mathrm{eff}}(r) = \dfrac{1}{r^{2}\left(1+\chi\right)}\left(1 - \dfrac{2M}{r}\right).
\end{align}

To perform ray--tracing simulations, it is necessary to express the radial motion in terms of the azimuthal angle $\phi$. Using the relation $\mathrm{d}r/\mathrm{d}\phi = (\mathrm{d}r/\mathrm{d}\tau)/(\mathrm{d}\phi/\mathrm{d}\tau)$, we arrive at the orbital equation:
\begin{align}
\left(\dfrac{\mathrm{d}r}{\mathrm{d}\phi}\right)^{2} 
&= r^4 \left[ \dfrac{1}{b^2} - \dfrac{1}{r^{2}\left(1+\chi\right)}\left(1 - \dfrac{2M}{r}\right)\right].
\end{align}
By introducing the variable substitution $u \equiv 1/r$, the equation of motion is transformed into a dimensionless differential form suitable for numerical integration:
\begin{align}
\label{eq:orbit_u}
\dfrac{\mathrm{d}u}{\mathrm{d}\phi} = \sqrt{\dfrac{1}{b^{2}} - \dfrac{u^{2}}{1+\chi}(1 - 2Mu)}.
\end{align}

The boundary of the black hole shadow is determined by the unstable circular photon orbit, known as the photon sphere. Its radius $r_p$ is defined by the critical conditions for the effective potential:
\begin{align}
V_{\mathrm{eff}}(r_p) = \dfrac{1}{b_p^2} \quad\text{and}\quad \dfrac{\mathrm{d}V_{\mathrm{eff}}}{\mathrm{d}r}\bigg|_{r=r_p} = 0.
\end{align}
Solving these conditions yields the radius of the photon sphere $r_p$ and the critical impact parameter $b_p$ \cite{gralla2020lensing,li2021observational,fathi2023observational,he2022shadow,shi2024shadow}:
\begin{align}
r_p = 3M \quad\text{and}\quad b_p = 3\sqrt{3\left(1+\chi\right)}M.
\end{align}

\begin{figure}[t]
\centering
\subfloat[$\chi=0.1$]{
\includegraphics[width=0.48\textwidth]{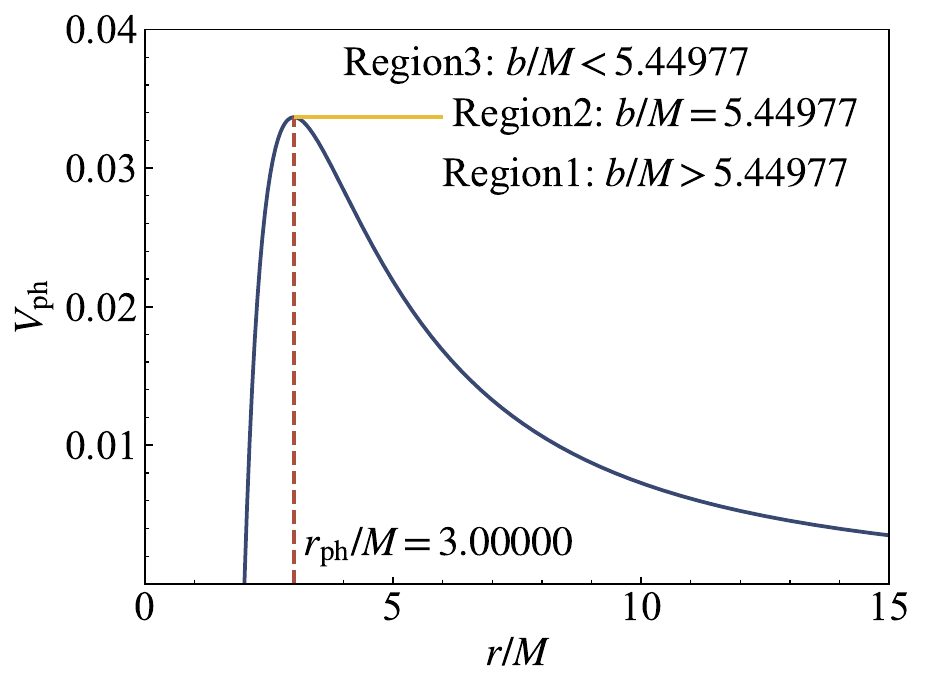}
}
\hfill
\subfloat[$\chi=0.3$]{
\includegraphics[width=0.48\textwidth]{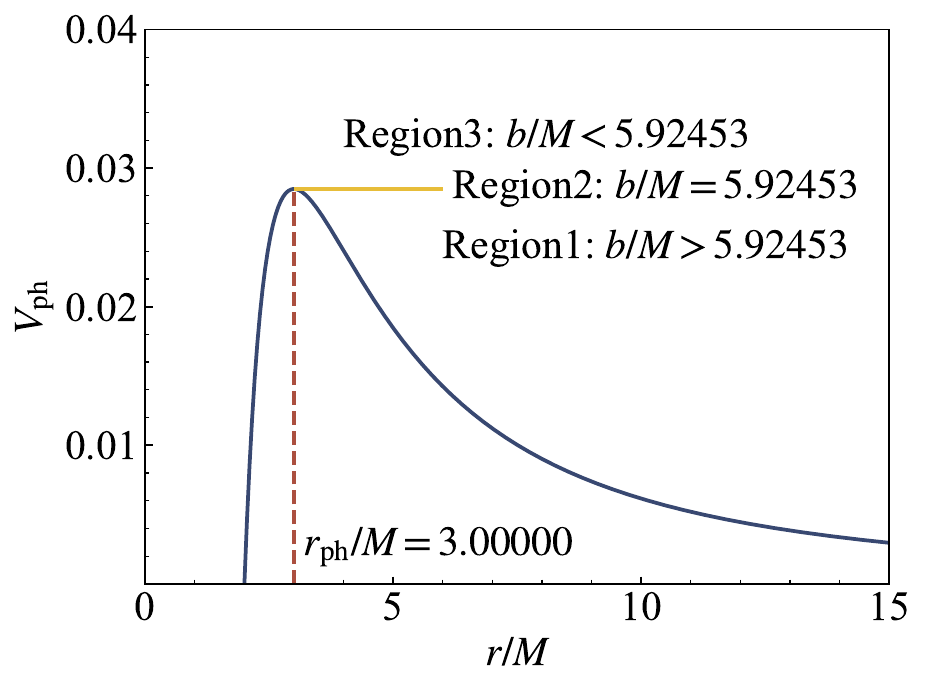}
}
\caption{The effective potential $V_{\mathrm{eff}}$ and the impact parameter $b_p$ of the new bumblebee black holes for different Lorentz violation parameters $\chi = 0.1$ and $\chi = 0.3$.}
\label{fig:Vph}
\end{figure}

To visualize the effects of Lorentz symmetry breaking on photon dynamics, we plot the effective potential $V_{\mathrm{eff}}(r)$ which has been defined in Eq.~\ref{Veff} for two representative cases, $\chi=0.1$ and $\chi=0.3$, in Fig.~\ref{fig:Vph}, respectively. These profiles quantitatively verify our analytical results. While the location of the potential barrier peak--corresponding to the photon sphere radius $r_{ph}$--remains fixed at $3M$ in both configurations, the critical impact parameter $b_p$ increases noticeably with $\chi$. Specifically, the critical value shifts from $b_p \approx 5.45M$ for $\chi=0.1$ to $b_p \approx 5.92M$ for $\chi=0.3$, indicating a magnification of the black hole shadow as the Lorentz violation parameter grows. 

Furthermore, Fig.~\ref{fig:Vph} explicitly delineate three distinct regimes of photon motion governed by the impact parameter $b$. These regimes correspond to the scattering region (Region 1, $b > b_p$) where light rays escape to infinity, the critical capture orbit (Region 2, $b = b_p$), and the capture region (Region 3, $b < b_p$) where photons plunge into the event horizon. This geometric modification, particularly the expansion of the capture region and the scaling of the shadow boundary, plays an essential role in determining the optical appearance of the accretion of matter surrounding the new bumblebee black hole, as it fundamentally alters the observable ring structures formed by photons emitted from the accretion disk.

\begin{figure}[t]
\centering
\subfloat[$\chi=0.1$\label{fig:traj1_1}]{
\includegraphics[width=0.48\textwidth]{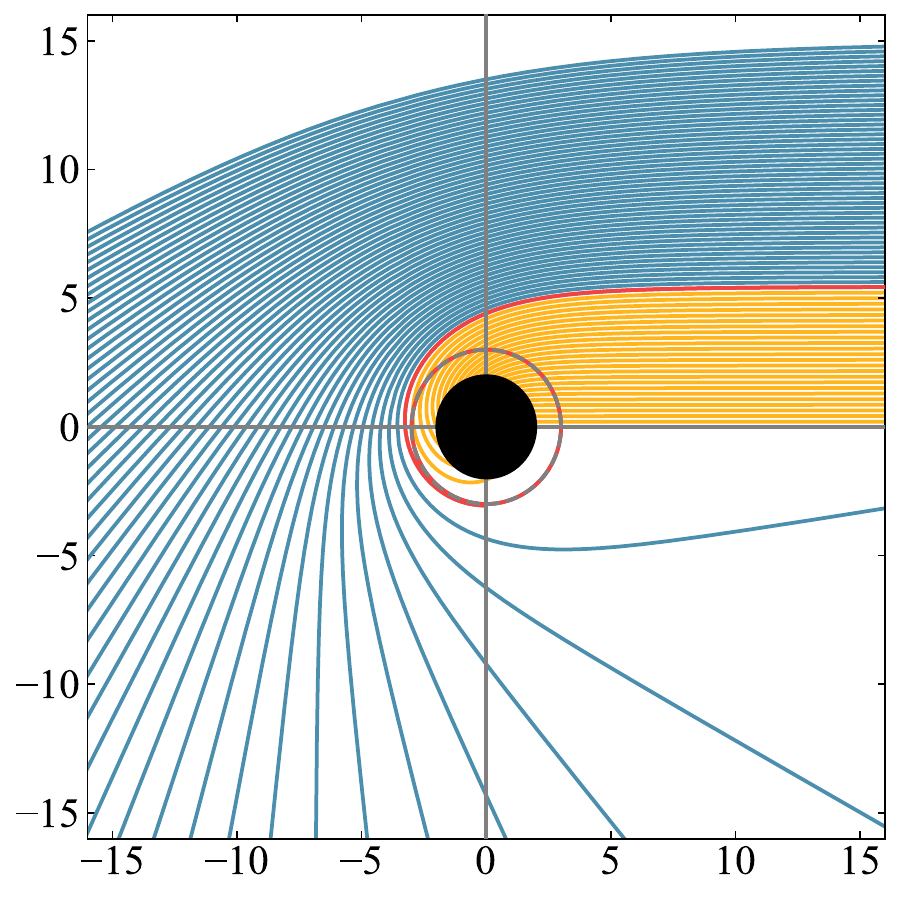}
}
\hfill
\subfloat[$\chi=0.3$\label{fig:traj1_2}]{
\includegraphics[width=0.48\textwidth]{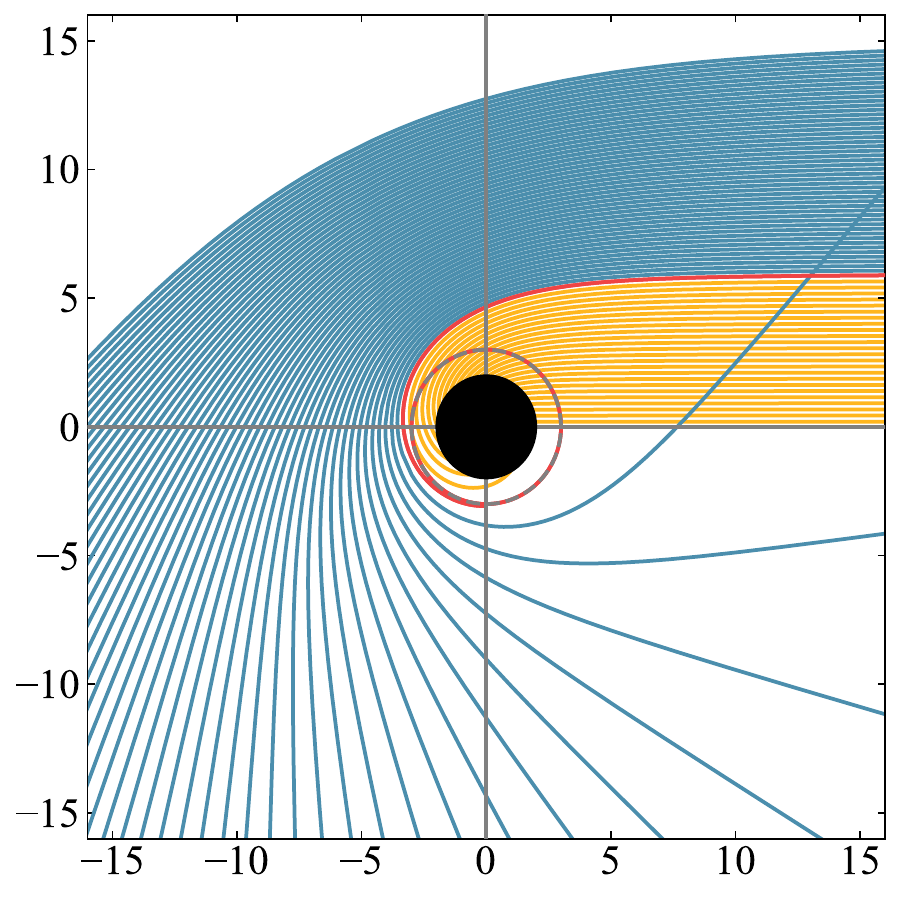}
}
\caption{Photon trajectories around the new bumblebee black hole in polar coordinates $(r, \phi)$ for different Lorentz violation parameters $\chi = 0.1$ and $\chi = 0.3$.}
\label{fig:traj1}
\end{figure}

To further visually corroborate the analytical results derived from the effective potential, we perform ray--tracing simulations to map the photon trajectories in the polar coordinates $(r, \phi)$. The results are presented in Fig.~\ref{fig:traj1}, corresponding to the Lorentz violation parameters $\chi=0.1$ and $\chi=0.3$, respectively. In these plots, the black disk represents the event horizon, while the grey dashed circle denotes the photon sphere at $r_{ph}=3M$.

The trajectories are color--coded to distinguish their physical fates, governed by the impact parameter $b$ relative to the critical value $b_p$. The blue curves represent scattering geodesics ($b > b_p$), where photons approach the black hole but are deflected by the potential barrier, eventually escaping to spatial infinity. The orange curves depict capture geodesics ($b < b_p$), where photons possess insufficient angular momentum to sustain an orbit, inevitably plunging into the event horizon. Separating these two regions is the red curve, which corresponds to the critical trajectory ($b = b_p$). As shown, photons on this path asymptotically approach the photon sphere, orbiting the black hole multiple times before instability dictates their final state.

A comparative analysis of Fig.~\ref{fig:traj1} provides remarkable evidence for the geometric modification induced by the bumblebee field. Although the physical location of the photon sphere remains invariant, the range of impact parameters leading to photon capture is clearly larger in the $\chi=0.3$ case. This expansion is visually demonstrated by the wider orange region in Fig.~\ref{fig:traj1_2} compared to Fig.~\ref{fig:traj1_1}. Such an observation confirms that the critical impact parameter $b_p$ shifts outward as $\chi$ increases. Consequently, light rays that would otherwise escape in a low-$\chi$ geometry are captured in the high-$\chi$ regime. This alteration in photon trajectories dictates how the specific intensity profiles from an accretion disk are mapped to the observer's image plane. Therefore, it influences the positions and widths of the observable direct emission, lensing rings, and photon rings, which we shall investigate in the following section.


\section{Shadows and rings of thin disk emission}

Black holes are seldom isolated vacuum solutions in practical astrophysical situations; instead, they are usually surrounded by bright matter, including accretion discs, which are the main sources of light for observation. Finding possible observational indicators of changed gravity therefore requires examining the optical appearance of a black hole lit by an accretion flow. In this section, we examine an optically and geometrically narrow accretion disc situated on the bumblebee black hole's equatorial plane. We examine the optical rings and shadows produced in this spacetime using the ray--tracing approach. We will systematically classify the photon trajectories based on their orbital behavior, derive the observed specific intensities using transfer functions, and finally simulate the observational appearances under various emission profiles to reveal the specific effects of the Lorentz violation parameter $\chi$.


\subsection{Direct emission, Lensing ring and photon ring}

The visual appearance of a black hole surrounded by an optically and geometrically thin accretion disk is governed by the trajectories of photons emitted from the disk and their subsequent deflection by the gravitational field. As discussed in previous literature, the image observed by a distant observer can be decomposed into distinct components based on the total number of orbits $n = \phi/2\pi$ that light rays complete around the central object \cite{gralla2019black,gralla2020lensing}.

Depending on the value of $n$, the trajectories are classified into three categories \cite{gralla2019black,gralla2020lensing,zeng2020influence,zeng2020shadows,zeng2022shadows,zeng2025holographic,li2021observational,li2021shadows}:
\begin{enumerate}
\item \textbf{Direct emission ($n < 3/4$)}: These light rays intersect the accretion disk only once. They constitute the primary image of the disk.
\item \textbf{Lensing ring ($3/4 < n < 5/4$)}: These photons traverse the equatorial plane twice, effectively capturing the emission from the back side of the disk.
\item \textbf{Photon ring ($n > 5/4$)}: These rays orbit the black hole at least three times, originating from the region asymptotically close to the photon sphere.
\end{enumerate}

\begin{figure}[t]
\centering
\includegraphics[width=0.48\textwidth]{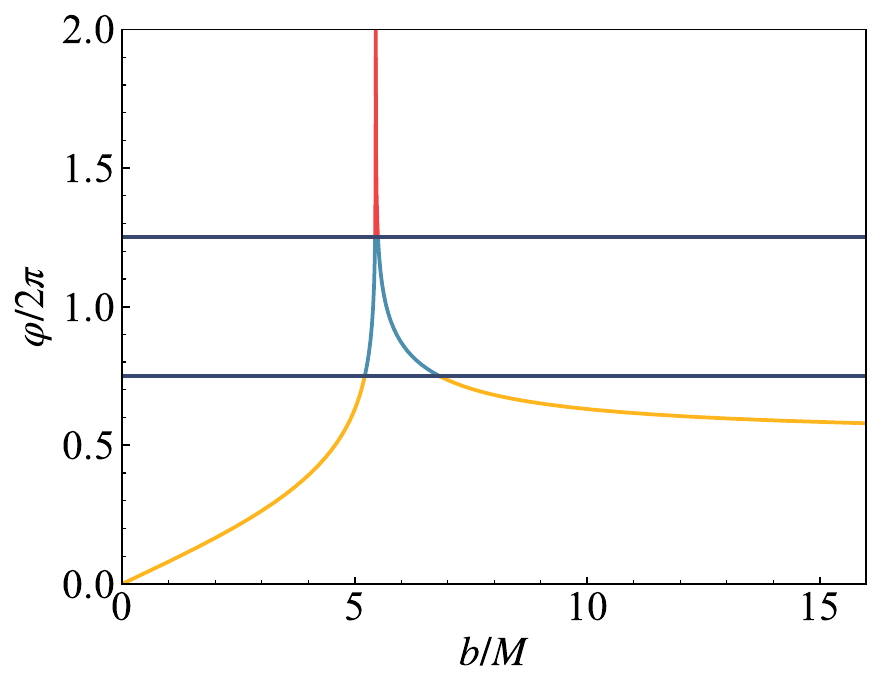}
\hfill
\includegraphics[width=0.48\textwidth]{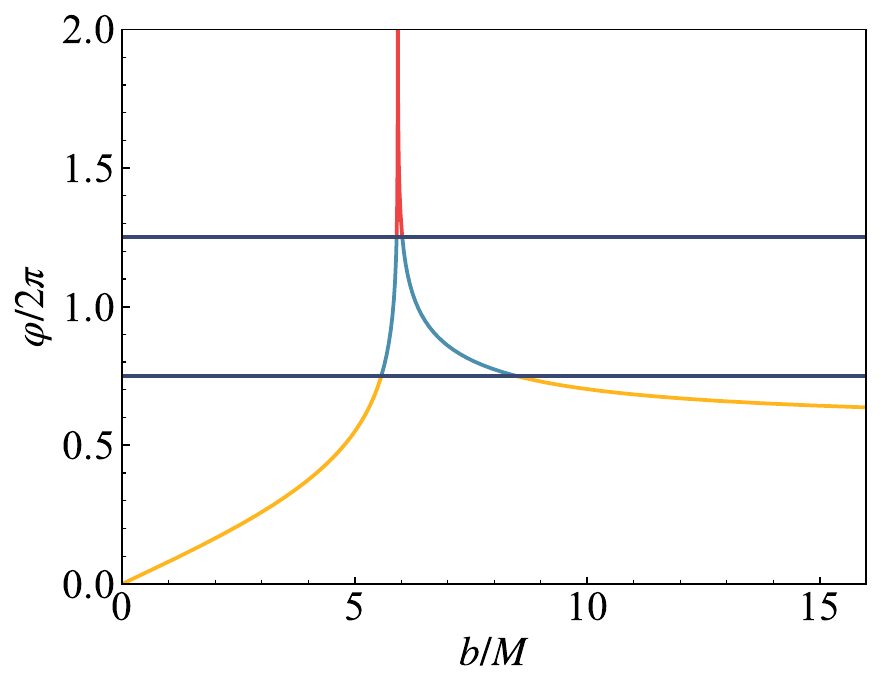}
\\
\subfloat[$\chi=0.1$\label{fig:traj2_1}]{
\includegraphics[width=0.48\textwidth]{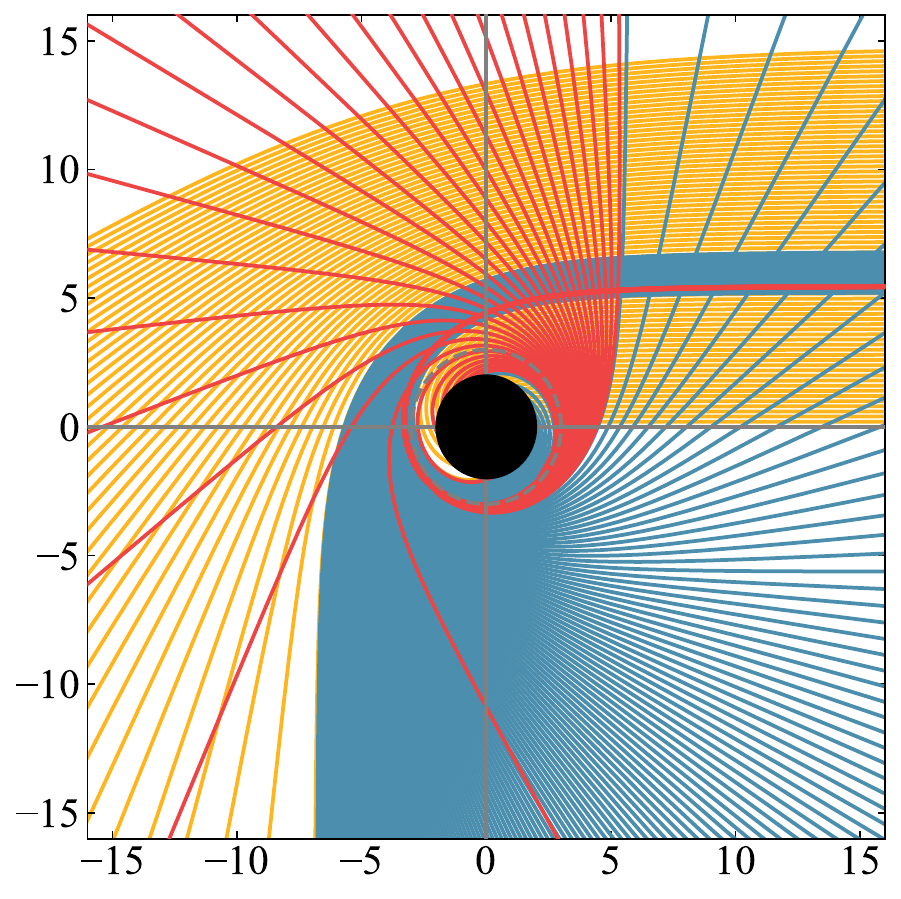}
}
\hfill
\subfloat[$\chi=0.3$\label{fig:traj2_2}]{
\includegraphics[width=0.48\textwidth]{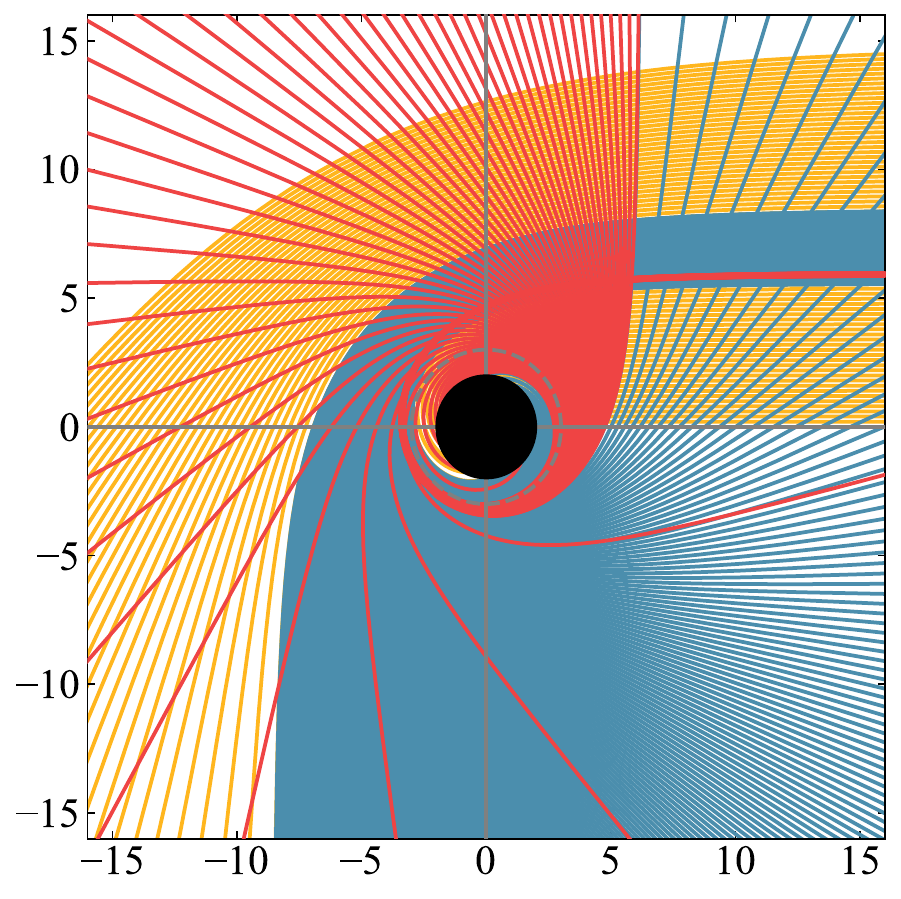}
}
\caption{Optical characteristics of the bumblebee black hole with Lorentz violation parameters $\chi = 0.1$ (left column) and $\chi = 0.3$ (right column).}
\label{fig:traj2}
\end{figure}

In the context of the new bumblebee black hole, the impact parameter ranges defining these regions are significantly modified by the Lorentz violation parameter $\chi$. To quantify this effect, we numerically calculate the orbital number $n$ as a function of the impact parameter $b$, as shown in the top row of Fig.~\ref{fig:traj2}. The vertical asymptotes in these plots indicate the positions of the critical impact parameter $b_p$, where the number of orbits diverges.

Utilizing the ray-tracing code, we determine the precise boundaries for the optical regions. For the representative case of $\chi = 0.1$, the impact parameter ranges for the direct emission, lensing ring, and photon ring are obtained as follows:
\begin{align}
\label{regions_chi0.1}
\begin{cases}
\text{Direct emission}: & b < 5.21654M \quad \text{and} \quad b > 6.81693M \\
\text{Lensing ring}: & 5.21654M < b < 5.43718M \quad \text{and} \quad 5.49812M < b < 6.81693M \\
\text{Photon ring}: & 5.43718M < b < 5.49812M
\end{cases}
\end{align}

With the increase of the Lorentz violation parameter to $\chi = 0.3$, these optical windows undergo a significant expansion and outward shift:
\begin{align}
\label{regions_chi0.3}
\begin{cases}
\text{Direct emission}: & b < 5.57097M \quad \text{and} \quad b > 8.47376M \\
\text{Lensing ring}: & 5.57097M < b < 5.89969M \quad \text{and} \quad 6.02185M < b < 8.47376M \\
\text{Photon ring}: & 5.89969M < b < 6.02185M
\end{cases}
\end{align}

The bottom row of Fig.~\ref{fig:traj2} shows the respective spatial paths of these light rays. Our numerical results are visually confirmed by comparing the left column ($\chi=0.1$) with the right column ($\chi=0.3$). The essential effect parameter $b_p$ moves outward from around $5.45M$ to $5.92M$ when $\chi$ grows. As a result, the visible rings--represented by the dense bundles of curved trajectories--appear at greater radial distances and the range of impact parameters related to photon capture grows. Additionally, the outer lensing ring window's width grows from around $1.32M$ to $2.45M$. This suggests that, in contrast to the Schwarzschild-like case, the higher-order optical rings are both enlarged and dispersed over a larger angular region on the observer's screen for a bumblebee black hole with a large $\chi$.

\subsection{Observed specific intensities and transfer functions}

Having classified the light trajectories, we now evaluate the observed specific intensity of the accretion disk. We assume the disk emits isotropically in the rest frame of static world-lines. According to Liouville's theorem, the ratio of the specific intensity to the cube of the frequency, $I_{\nu}/\nu^3$, is invariant along a light ray. Therefore, the specific intensity observed by a distant observer at frequency $\nu_{obs}$, $I_{obs}(\nu_{obs})$, is related to the emitted specific intensity $I_{emi}(\nu_{emi})$ at the disk by \cite{zeng2020influence,zeng2020shadows,zeng2022shadows,zeng2025holographic,li2021observational,li2021shadows}:
\begin{align}
I_{obs}(\nu_{obs}) = g^3 I_{emi}(\nu_{emi}),
\end{align}
where $g = \nu_{obs}/\nu_{emi} = \sqrt{-g_{tt}}$ is the redshift factor. For the bumblebee black hole metric Eq.~\ref{metric} considered here, the redshift factor is given by:
\begin{align}
g = \sqrt{\dfrac{1}{1+\chi}\left(1 - \dfrac{2M}{r}\right)}.
\end{align}
Integrating over all frequencies, the total observed intensity $I_{obs}$ is related to the total emitted intensity $I_{emi}(r)$. Since the accretion disk is located in the equatorial plane, a single light ray can intersect the disk multiple times, picking up additional intensity at each crossing. The total observed intensity for a given impact parameter $b$ is therefore the sum of intensities from all intersections \cite{meng2023images}:
\begin{align}
I_{obs}(b) = \sum_{n} \left[\dfrac{1}{1+\chi}\left(1 - \dfrac{2M}{r}\right)\right]^{2} I_{emi}(r)|_{r=r_n(b)},
\end{align}
where $r_n(b)$ is the \textit{transfer function}, representing the radial coordinate of the $n$--th intersection of the light ray with the accretion disk.

\begin{figure}[t]
\centering
\subfloat[$\chi=0.1$\label{fig:transfer_1}]{
\includegraphics[width=0.48\textwidth]{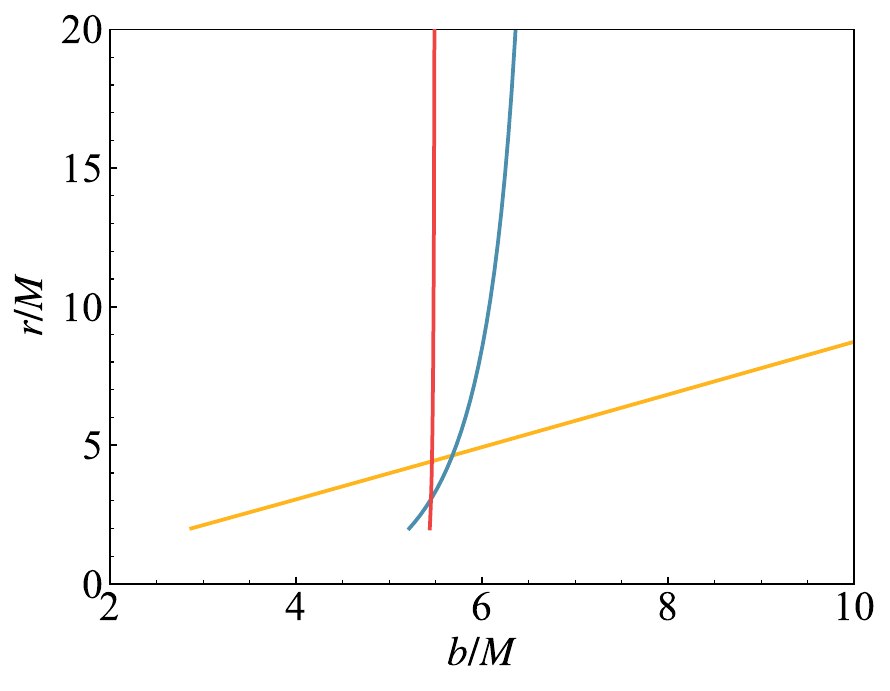}
}
\hfill
\subfloat[$\chi=0.3$\label{fig:transfer_2}]{
\includegraphics[width=0.48\textwidth]{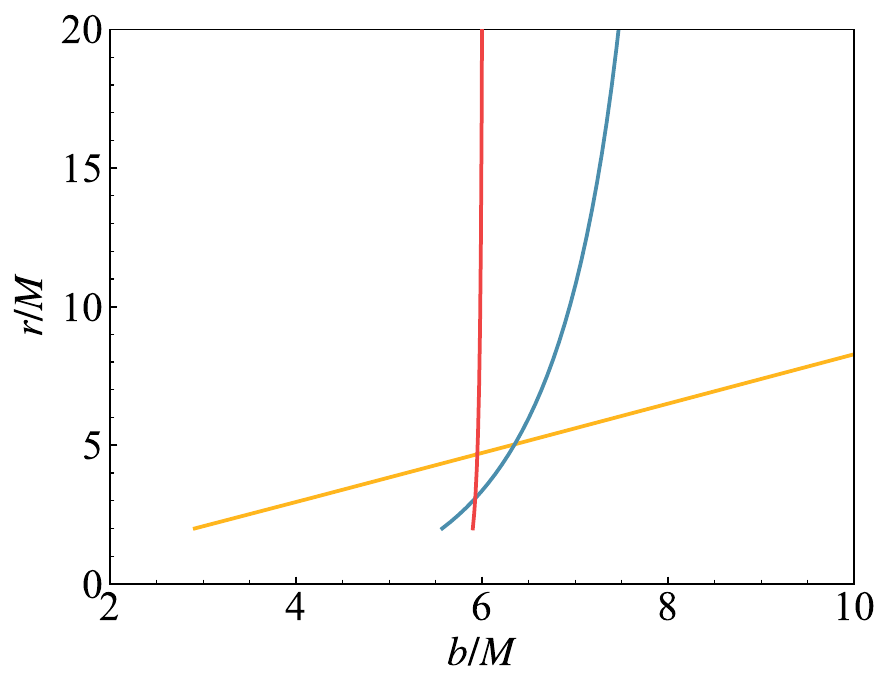}
}
\caption{The first three transfer functions $r_{n}(b)$ for the bumblebee black hole with $\chi=0.1$ and $\chi=0.3$. The yellow, blue, and red curves correspond to the direct emission ($n=1$), lensing ring ($n=2$), and photon ring ($n=3$), respectively.}
\label{fig:transfer}
\end{figure}

The behavior of the transfer functions for the first three image orders ($n=1, 2, 3$) is plotted in Fig.~\ref{fig:transfer}. The yellow curve corresponds to the direct image ($n=1$), which is a linear--like mapping with a slope $\mathrm{d}r/\mathrm{d}b \approx 1$ \cite{gralla2019black,li2021observational,fathi2023observational,he2022shadow,shi2024shadow,zeng2020influence}. This indicates that the direct image is a relatively faithful reproduction of the disk's emission profile. The blue curve represents the lensing ring ($n=2$), where the slope $\mathrm{d}r/\mathrm{d}b$ increases rapidly, implying that a large radial section of the disk's back side is mapped onto a narrow range of impact parameters. The red curve corresponds to the photon ring ($n=3$); its slope approaches infinity, indicating extreme de-magnification where the entire front side of the disk is compressed into an infinitesimally thin ring. Comparing Fig.~\ref{fig:transfer_1} and Fig.~\ref{fig:transfer_2}, we observe that while the qualitative behavior of the transfer functions remains similar, the critical impact parameter where the slopes diverge shifts to higher values for larger $\chi$, consistent with our previous findings.

\subsection{Observational appearances of direct emissions and rings}

With the transfer functions established, we can now compute the final observational appearance of the new bumblebee black hole by convolving the transfer functions with specific emission profiles $I_{emi}(r)$. Since the intrinsic luminosity distribution of accretion disks depends on complex astrophysical processes, we employ three representative toy models to systematically investigate the optical signatures \cite{wang2022optical,yang2023shadow,li2021shadows,zeng2020influence,zeng2020shadows,zeng2022shadows,zeng2025holographic,meng2023images,gan2021photon,guo2022gravitational,chen2022appearance}. These models feature different emission starting points and decay rates, allowing us to disentangle the geometric effects of the spacetime from the source morphology.


\subsubsection{Emission starting from the ISCO}

In the first model, we consider an accretion disk whose emission is sharply truncated at the inner edge, corresponding to the Innermost Stable Circular Orbit (ISCO), and decays rapidly with radial distance. For the static spherically symmetric metric, the ISCO radius is located at $r_{isco} = 6M$, identical to the Schwarzschild case and independent of the parameter $\chi$. The emitted specific intensity is modeled by a second--order decay function:
\begin{align}
\label{model_2nd}
I_{emi}(r) = 
\begin{cases} 
\dfrac{1}{[r - (r_{isco} - 1)]^2} & r > r_{isco} \\
0 & r \leq r_{isco}
\end{cases}.
\end{align}

\begin{figure}[t]
\centering
\subfloat[$\chi=0.1$]{
\includegraphics[width=0.32\textwidth]{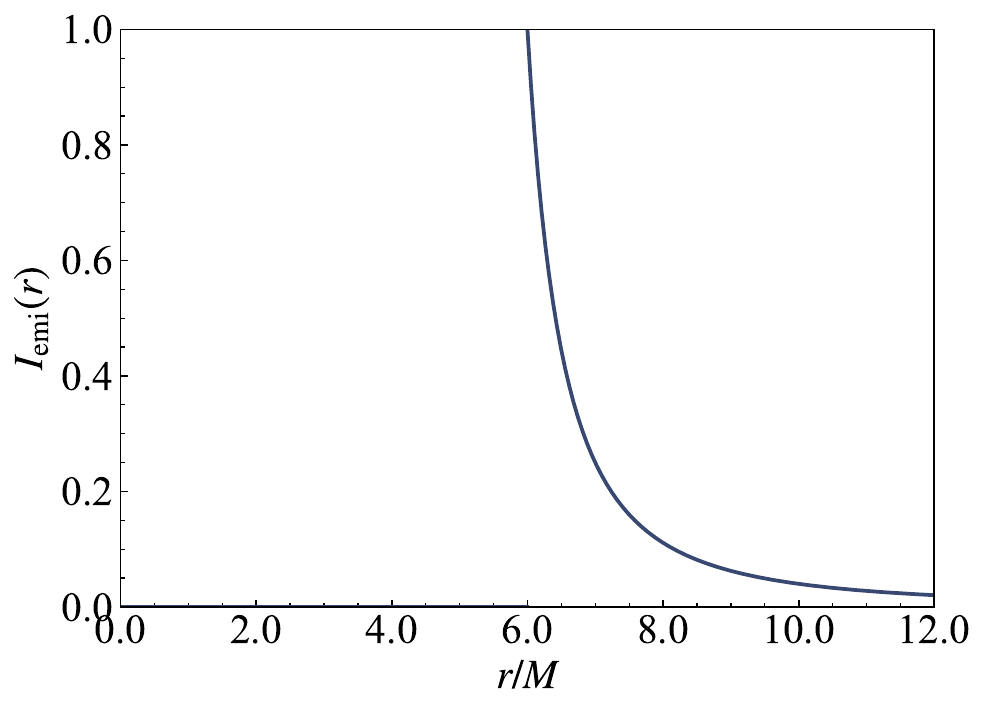}
\hfill
\includegraphics[width=0.32\textwidth]{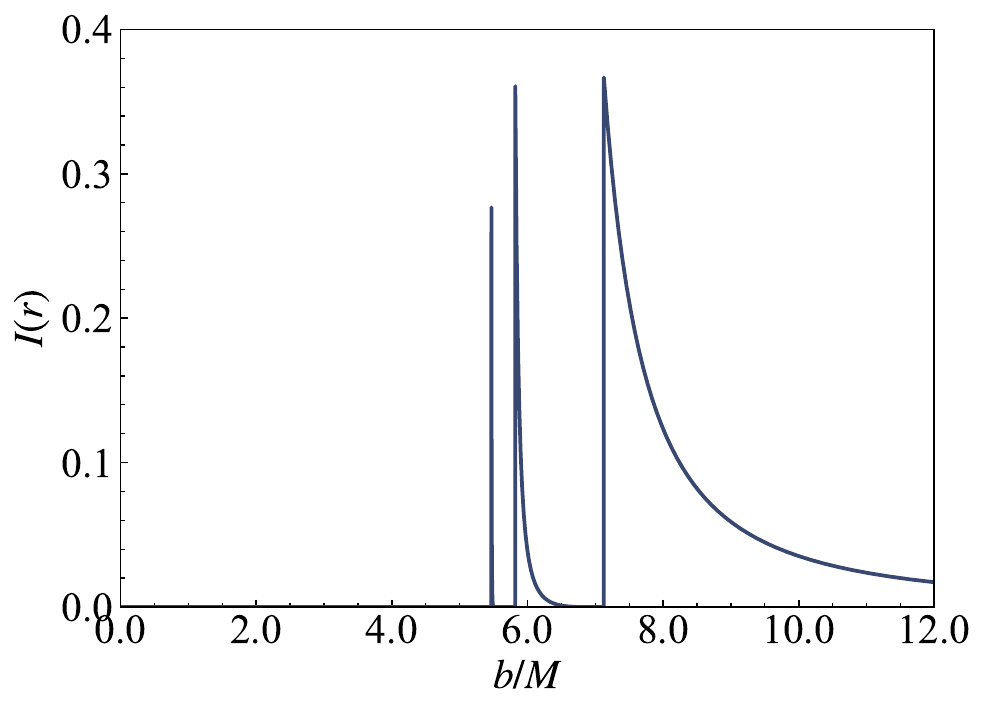}
\hfill
\includegraphics[width=0.32\textwidth]{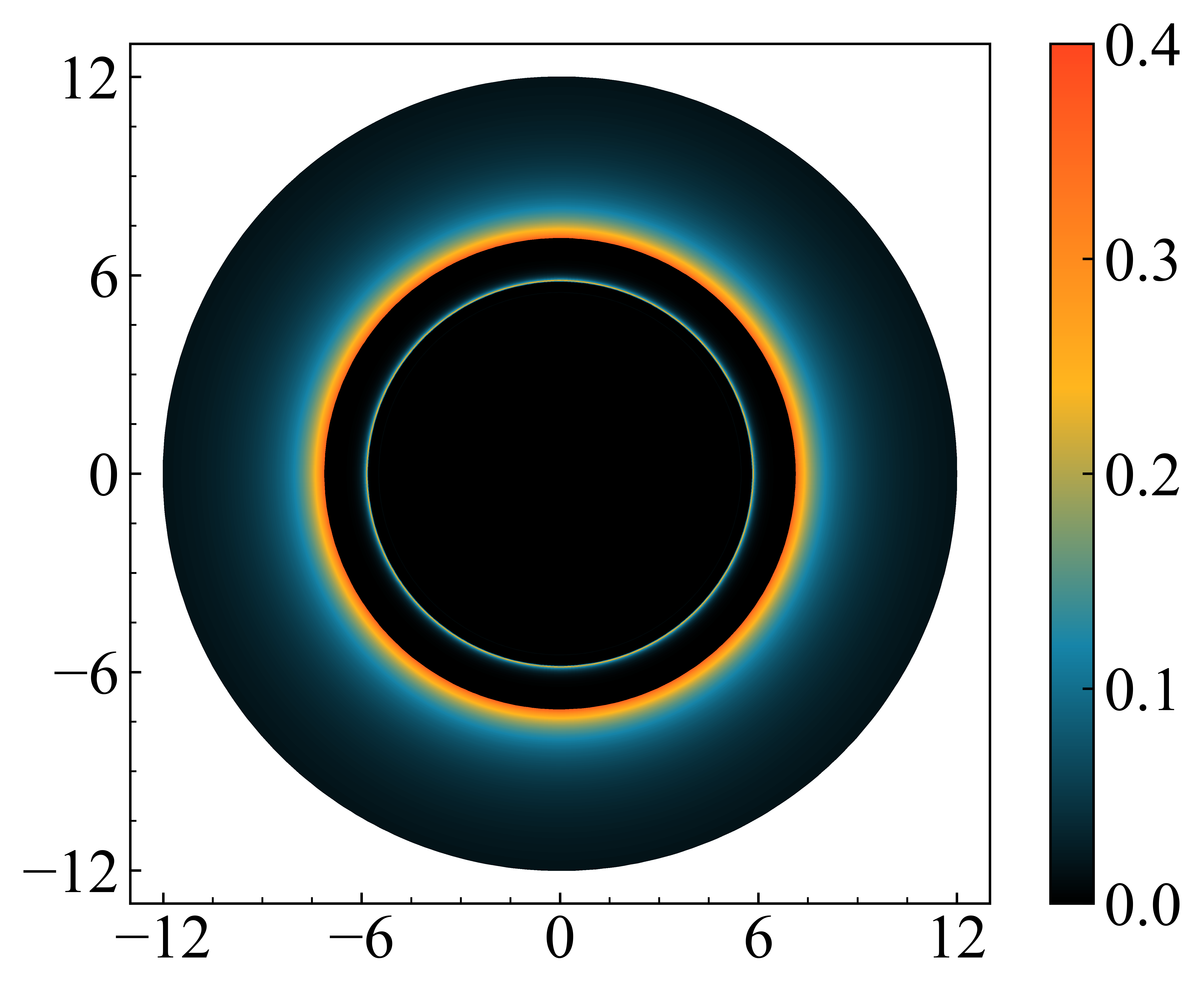}
}
\\
\subfloat[$\chi=0.3$]{
\includegraphics[width=0.32\textwidth]{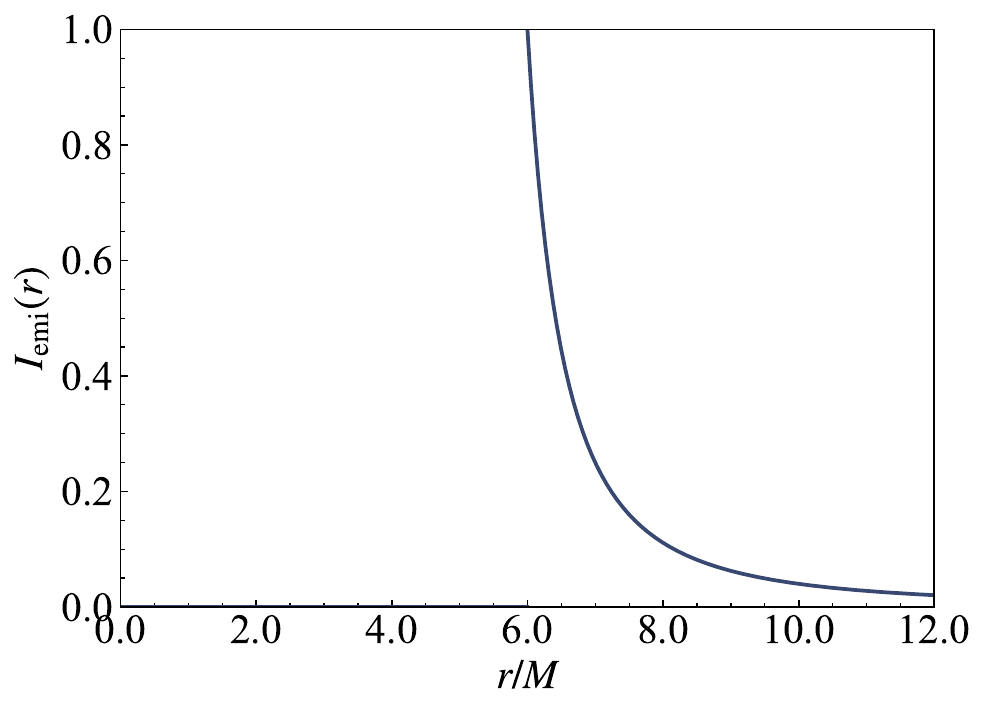}
\hfill
\includegraphics[width=0.32\textwidth]{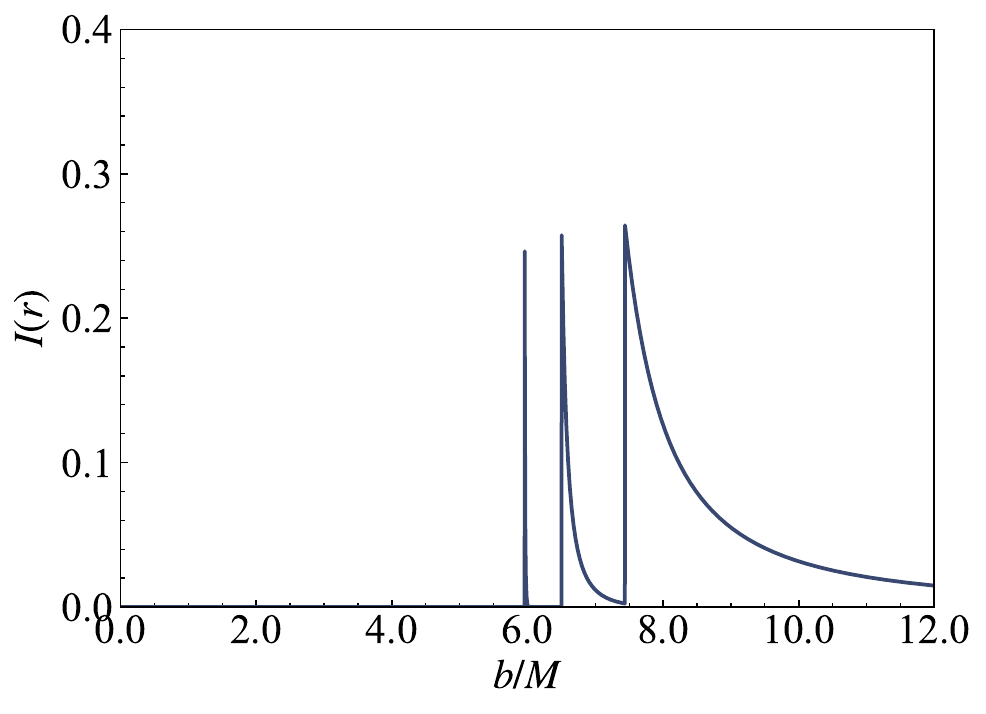}
\hfill
\includegraphics[width=0.32\textwidth]{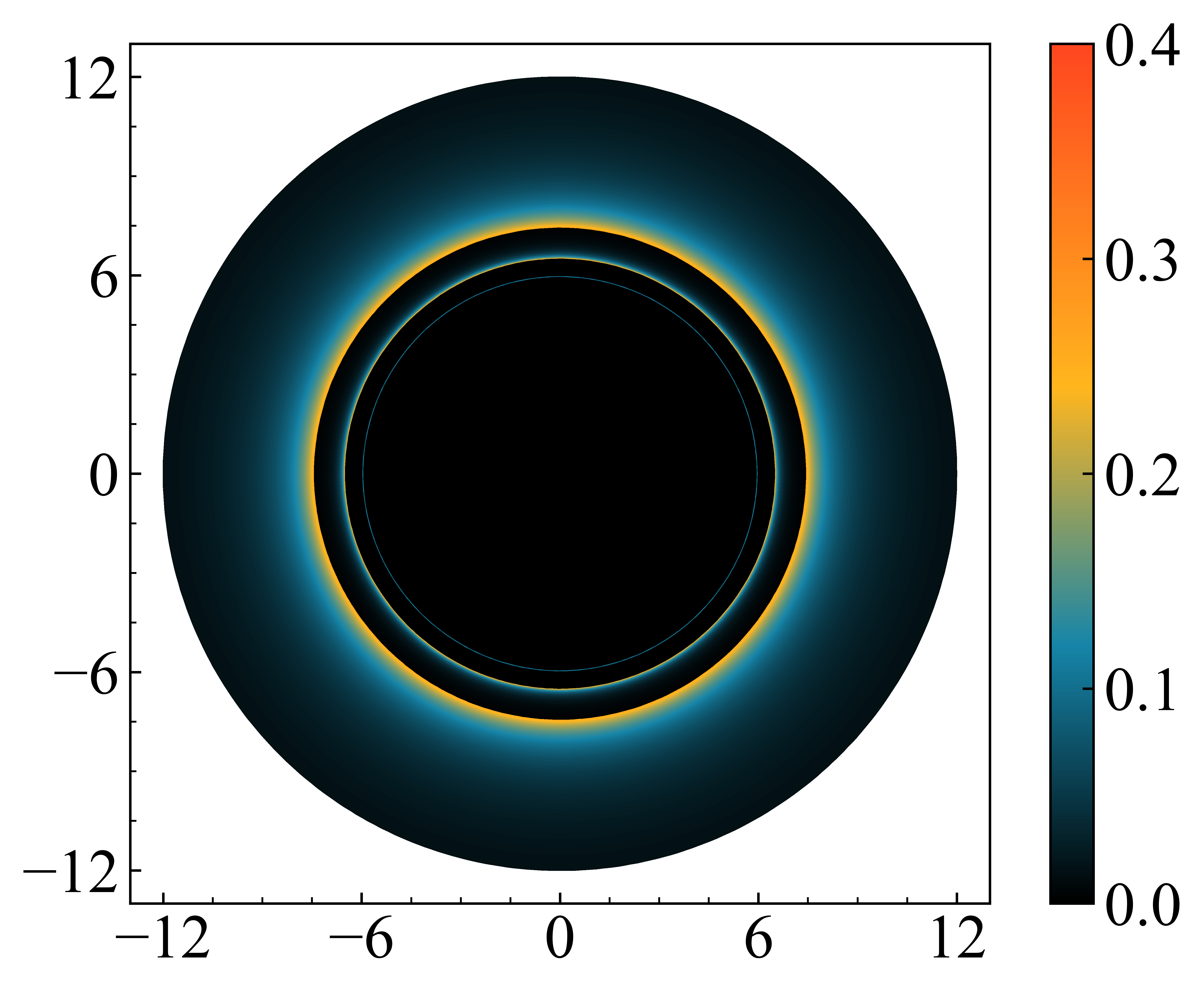}
}
\caption{Observational appearances of the bumblebee black hole for the second-order decay function.}
\label{fig:Iemi_2nd}
\end{figure}

The resulting optical signatures are presented in Fig.~\ref{fig:Iemi_2nd}. As illustrated in the left column, the intrinsic emission profile $I_{emi}(r)$ peaks sharply at $r=6M$ and vanishes for $r < 6M$. The corresponding observed intensity $I_{obs}(b)$, shown in the middle column, reveals that the observational appearance is overwhelmingly dominated by the direct emission ($n=1$). The profile of this primary image closely mimics the source but is gravitationally redshifted and mapped to a specific range on the observer's screen. Crucially, the Lorentz violation parameter $\chi$ significantly affects both the position and the magnitude of the intensity peak. For the case of $\chi=0.1$, the maximum observed intensity is approximately $0.37$, located at an impact parameter of $b \approx 7.13M$. In contrast, for $\chi=0.3$, the peak intensity decreases to approximately $0.26$ and shifts outward to $b \approx 7.44M$. This quantitative behavior indicates that a stronger bumblebee field not only expands the apparent size of the accretion disk image but also dilutes the observed flux due to the redistribution of light over a larger solid angle.

Regarding the fine structure of the image, the lensing ring ($n=2$) appears as a distinct, narrow spike interior to the direct image, while the photon ring ($n=3$) is extremely de-demagnified and manifests as a negligible feature near the critical curve. The right column visualizes these features in the shadow images. A comparison between the top row ($\chi=0.1$) and the bottom row ($\chi=0.3$) clearly demonstrates the geometric expansion effect induced by Lorentz symmetry breaking. As $\chi$ increases, the central dark area (the shadow) enlarges, and the bright rings move to larger radii, confirming that the parameter $\chi$ acts as a global scaling factor that governs the macroscopic scale of the observational features.


\subsubsection{Emission starting from the photon sphere}

In the second model, we investigate an accretion scenario where the emission extends further inward, originating directly from the photon sphere $r_{ph}$. This setup is particularly relevant for studying the contribution of relativistic images, as the source luminosity is concentrated in the region of strongest gravitational lensing. The emitted specific intensity is described by a third--order decay function:
\begin{align}
\label{model_3rd}
I_{emi}(r) = 
\begin{cases} 
\dfrac{1}{[r - (r_{ph} - 1)]^3} & r > r_{ph}\\
0 & r \leq r_{ph}
\end{cases},
\end{align}
where $r_{ph} = 3M$ is the radius of the photon sphere, which remains invariant under the bumblebee gravity modification.

\begin{figure}[t]
\centering
\subfloat[$\chi=0.1$]{
\includegraphics[width=0.32\textwidth]{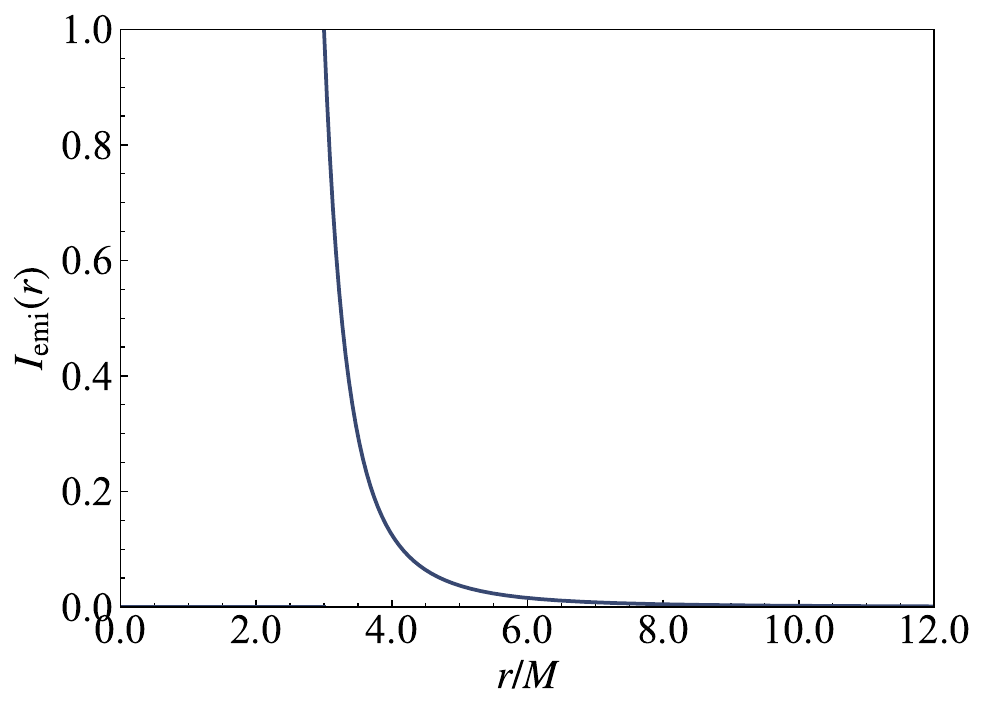}
\hfill
\includegraphics[width=0.32\textwidth]{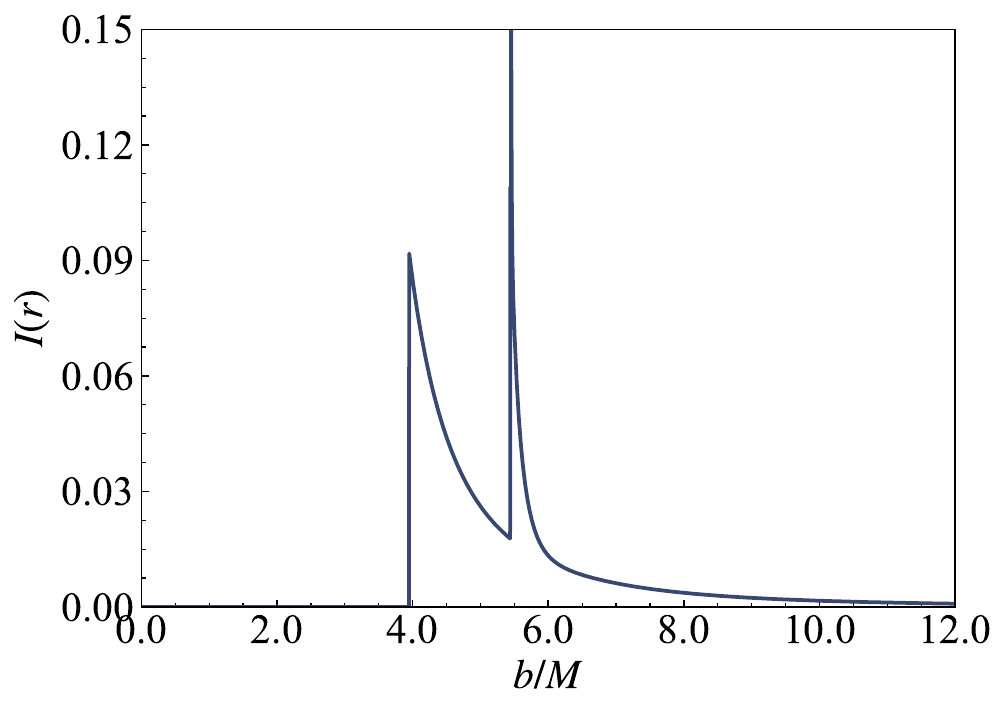}
\hfill
\includegraphics[width=0.32\textwidth]{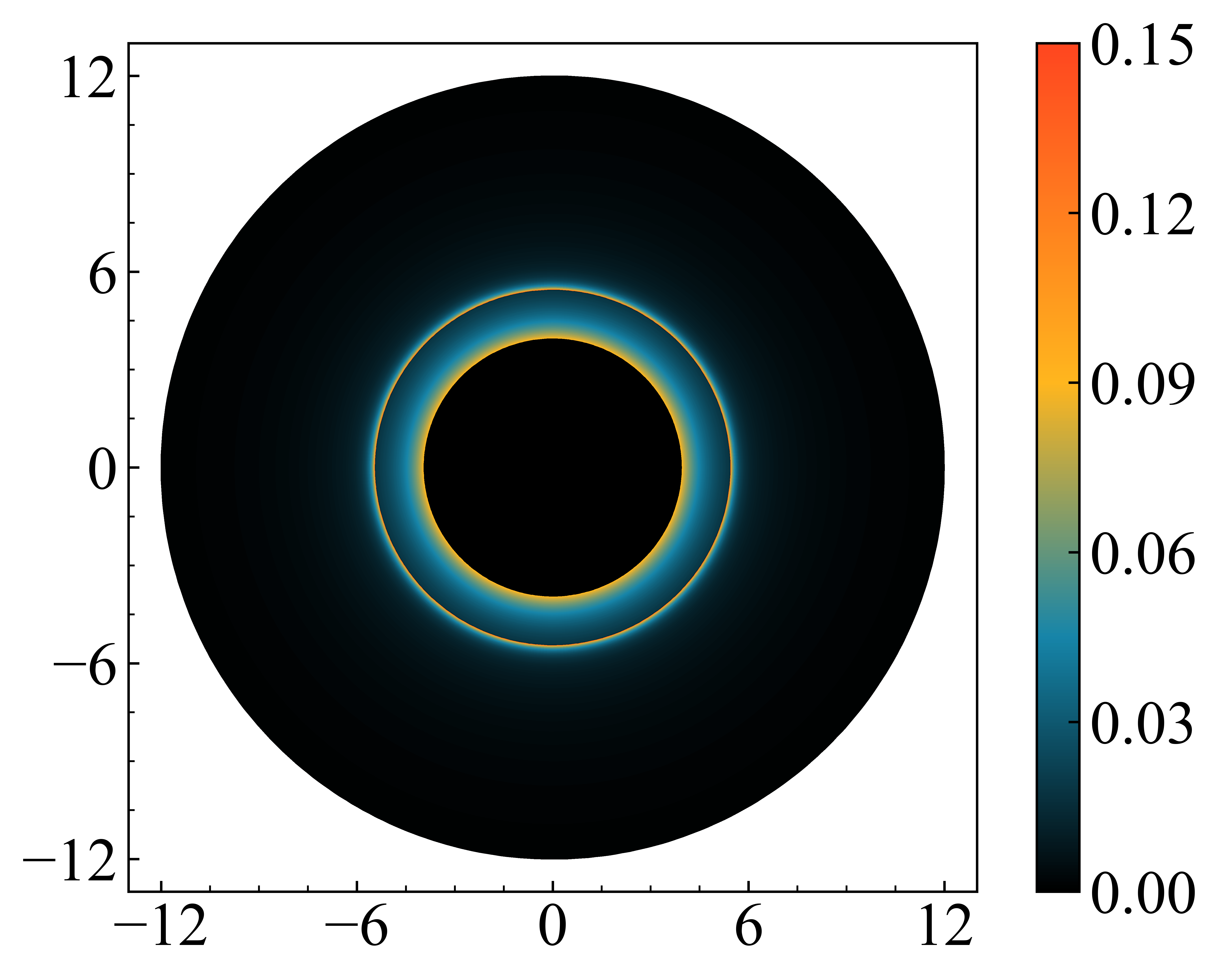}
}
\\
\subfloat[$\chi=0.3$]{
\includegraphics[width=0.32\textwidth]{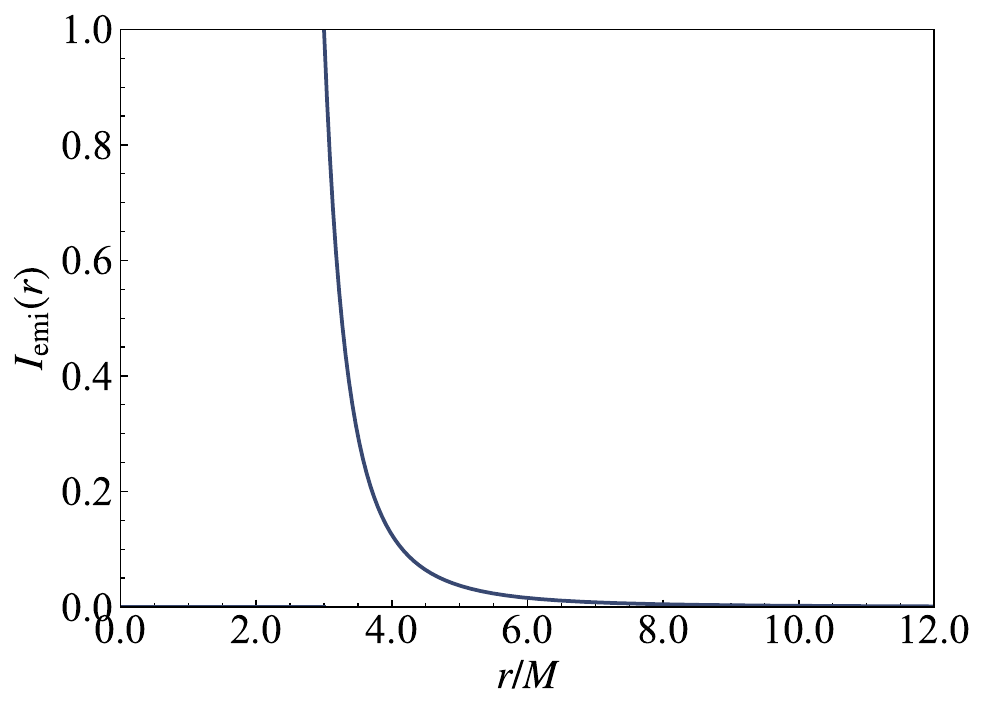}
\hfill
\includegraphics[width=0.32\textwidth]{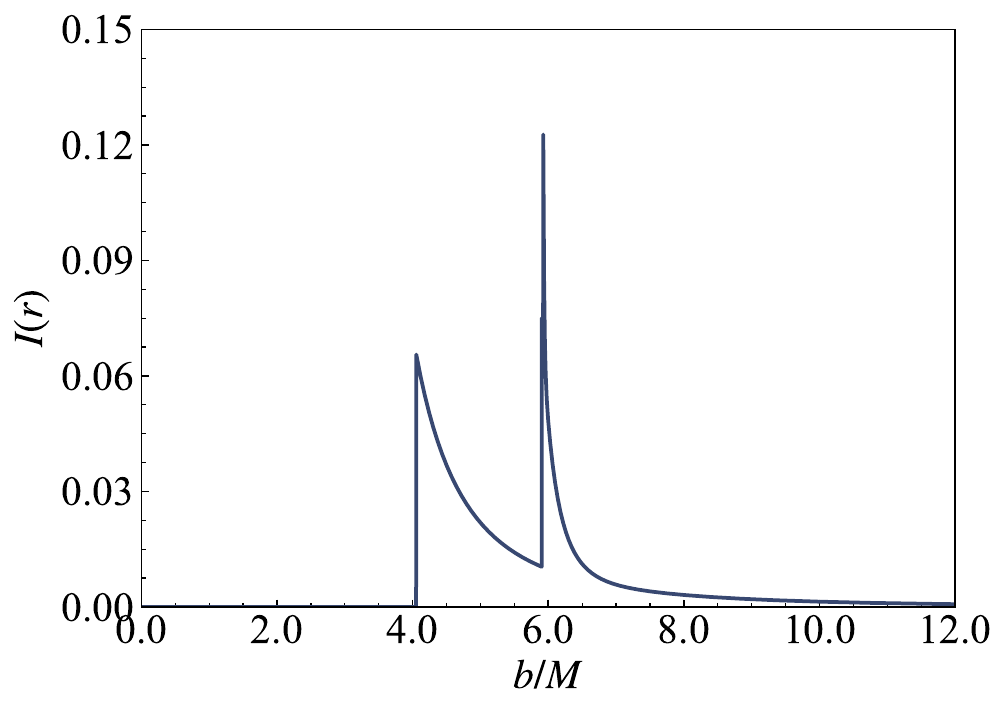}
\hfill
\includegraphics[width=0.32\textwidth]{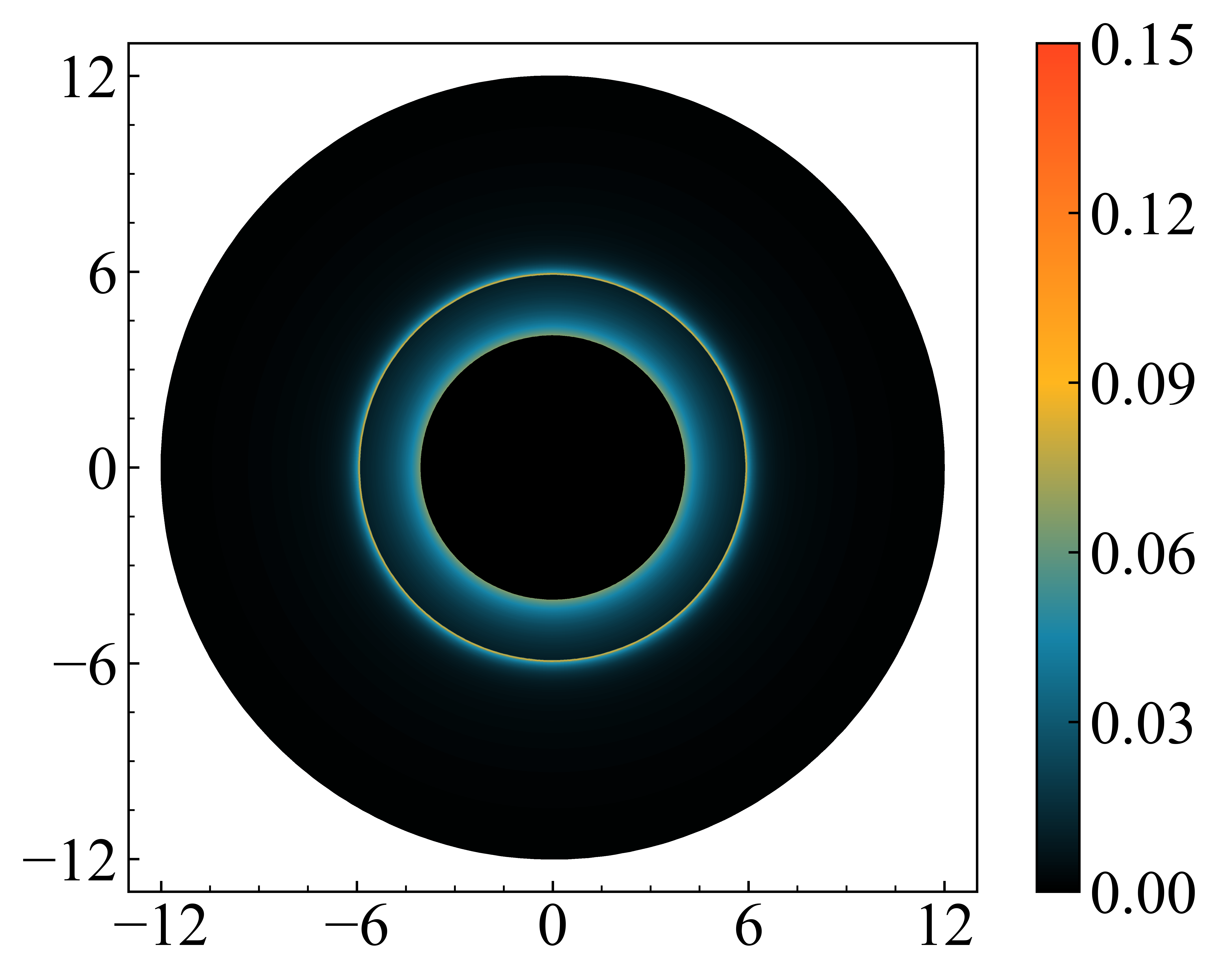}
}
\caption{Observational appearances of the bumblebee black hole for the third-order decay function.}
\label{fig:Iemi_3rd}
\end{figure}

The observational signatures for this model are displayed in Fig.~\ref{fig:Iemi_3rd}. The intrinsic emission profiles (left column) peak sharply at $r=3M$ and decay cubically with distance. Unlike the previous model where the rings were spatially separated, the extension of emission to the photon sphere leads to a significant superposition of the optical components. As shown in the observed intensity profiles (middle column), the lensing ring and photon ring are superimposed on the inner edge of the direct emission. This superposition creates a distinctive multi--peaked structure in the intensity profile, where the sharp spike corresponds to the combined flux of the photon and lensing rings effectively concentrated near the critical curve.

Quantitatively, the impact of Lorentz symmetry breaking is manifested in the position and prominence of this critical feature. For $\chi=0.1$, the observed intensity reaches its global maximum near the critical impact parameter $b_p \approx 5.45M$. However, as the Lorentz violation parameter increases to $\chi=0.3$, this peak shifts outward to $b \approx 5.92M$. This shift is consistent with the expansion of the shadow radius discussed earlier. Furthermore, a comparison of the images (right column) reveals that the prominent annular structure--resulting from the coalescence of the direct emission and relativistic rings--becomes noticeably wider and is located at a larger radius in the $\chi=0.3$ case. This demonstrates that for emissions originating near the photon sphere, the parameter $\chi$ plays a fundamental role in determining the angular size of the primary observable feature.


\subsubsection{Emission starting from the event horizon}

Finally, we consider a scenario where the accretion flow extends all the way to the event horizon $r_h$. This model represents a geometrically thin disk with a moderate decay rate, allowing us to probe the observational signatures of radiation emerging from the deepest regions of the potential well. The emitted specific intensity is given by a moderate decay function:
\begin{align}
\label{model_mod}
I_{emi}(r) = 
\begin{cases} 
\dfrac{\pi/2 - \arctan[r - (r_{isco} - 1)]}{\pi/2 - \arctan(r_p)} & r > r_h\\
0 & r \leq r_h
\end{cases},
\end{align}
where $r_h$ is the event horizon radius, which depends on the specific metric parameters in Eq.~\ref{metric}.

\begin{figure}[t]
\centering
\subfloat[$\chi=0.1$]{
\includegraphics[width=0.32\textwidth]{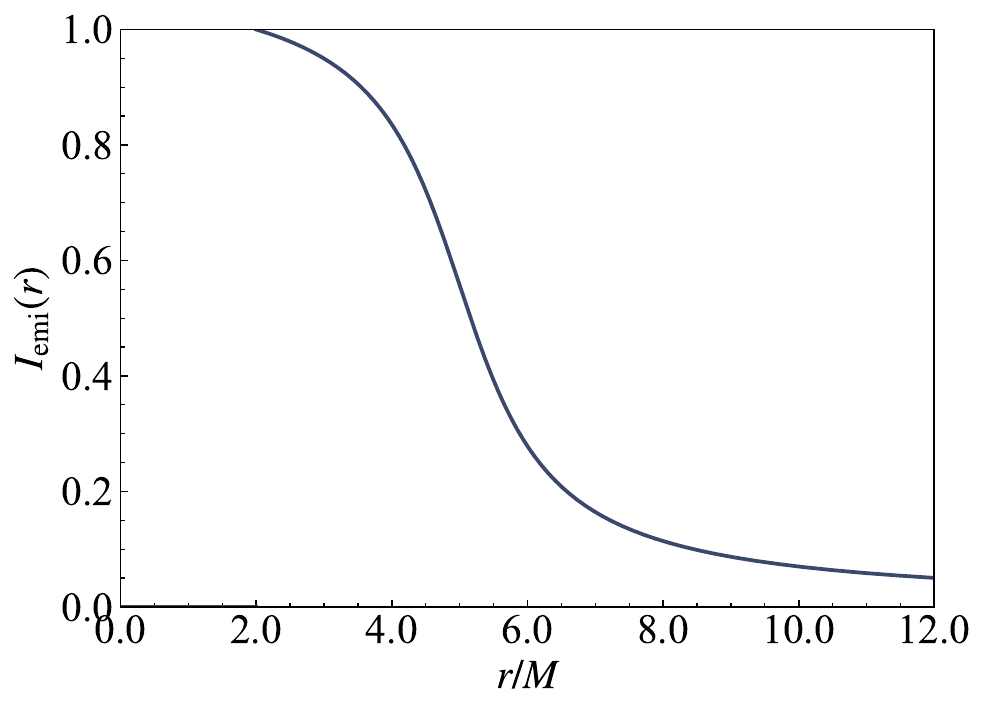}
\hfill
\includegraphics[width=0.32\textwidth]{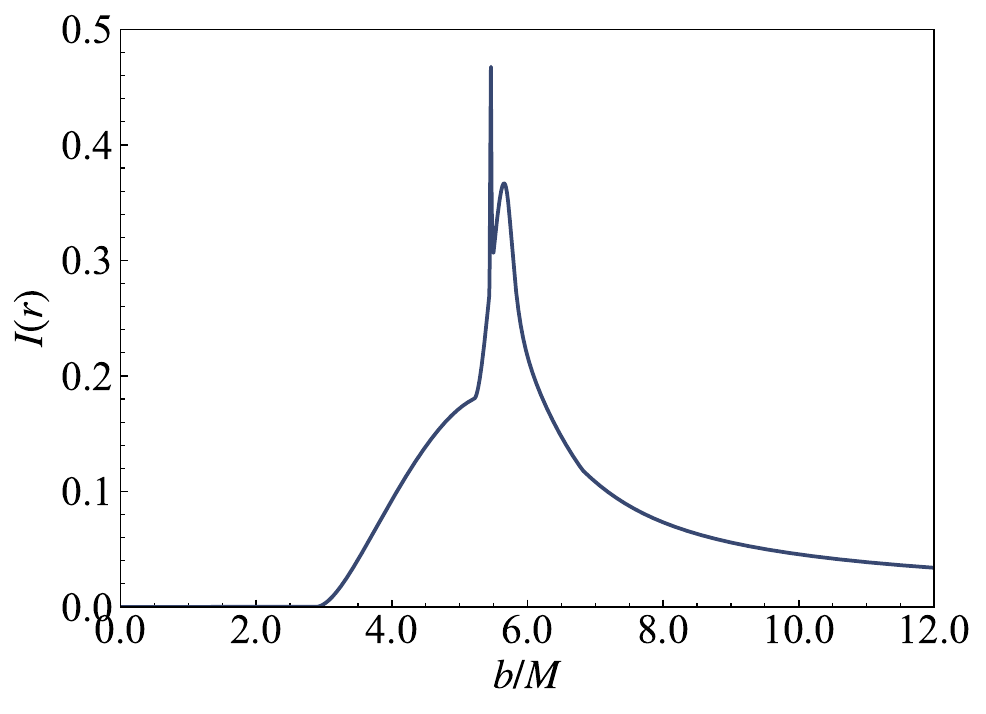}
\hfill
\includegraphics[width=0.32\textwidth]{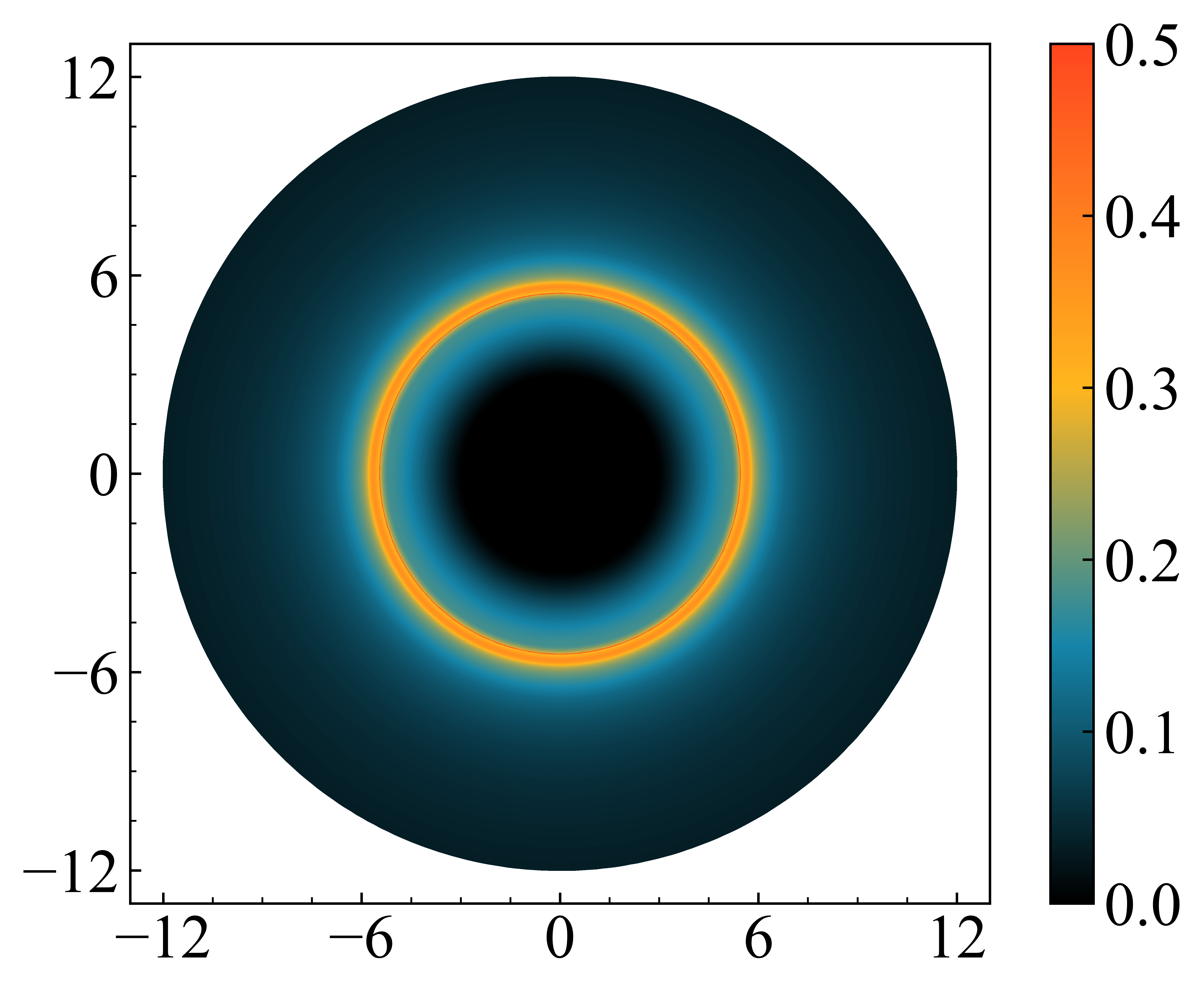}
}
\\
\subfloat[$\chi=0.3$]{
\includegraphics[width=0.32\textwidth]{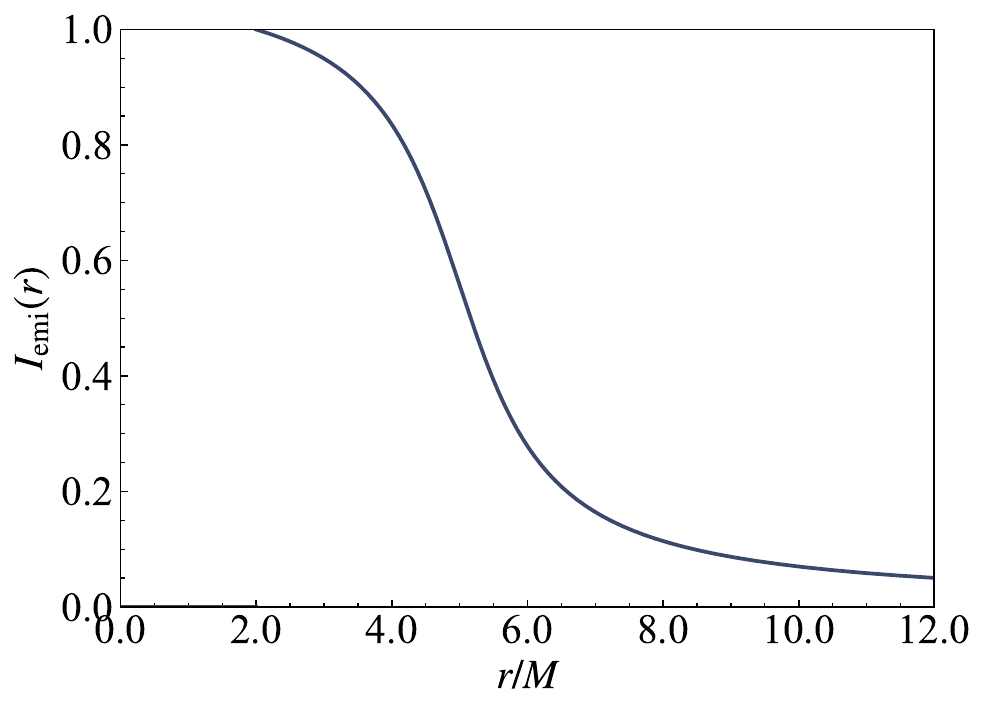}
\hfill
\includegraphics[width=0.32\textwidth]{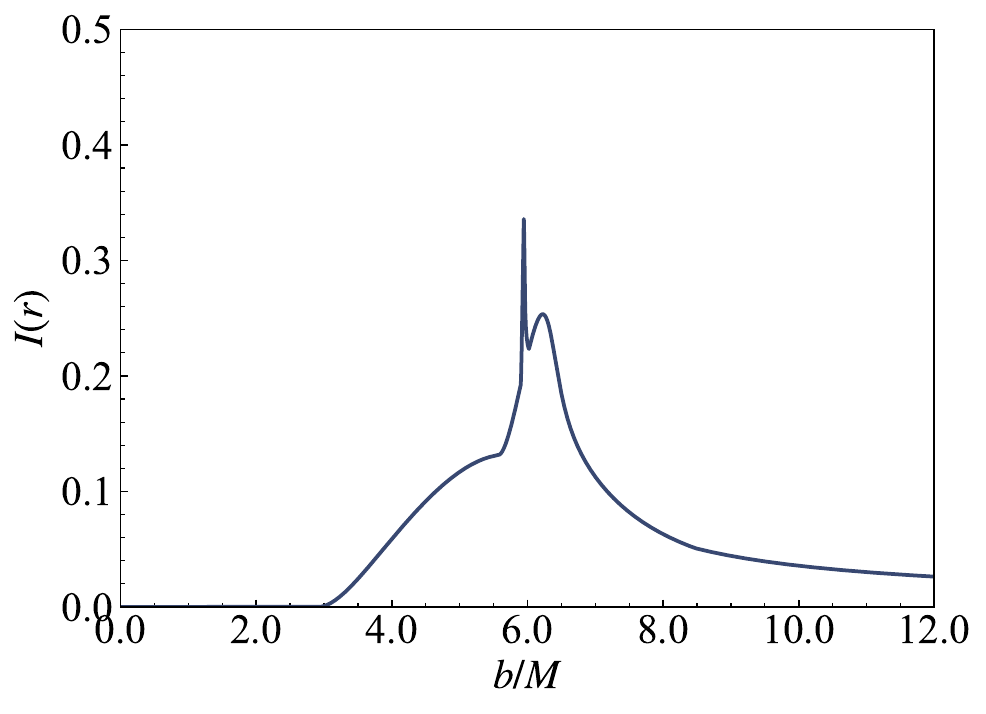}
\hfill
\includegraphics[width=0.32\textwidth]{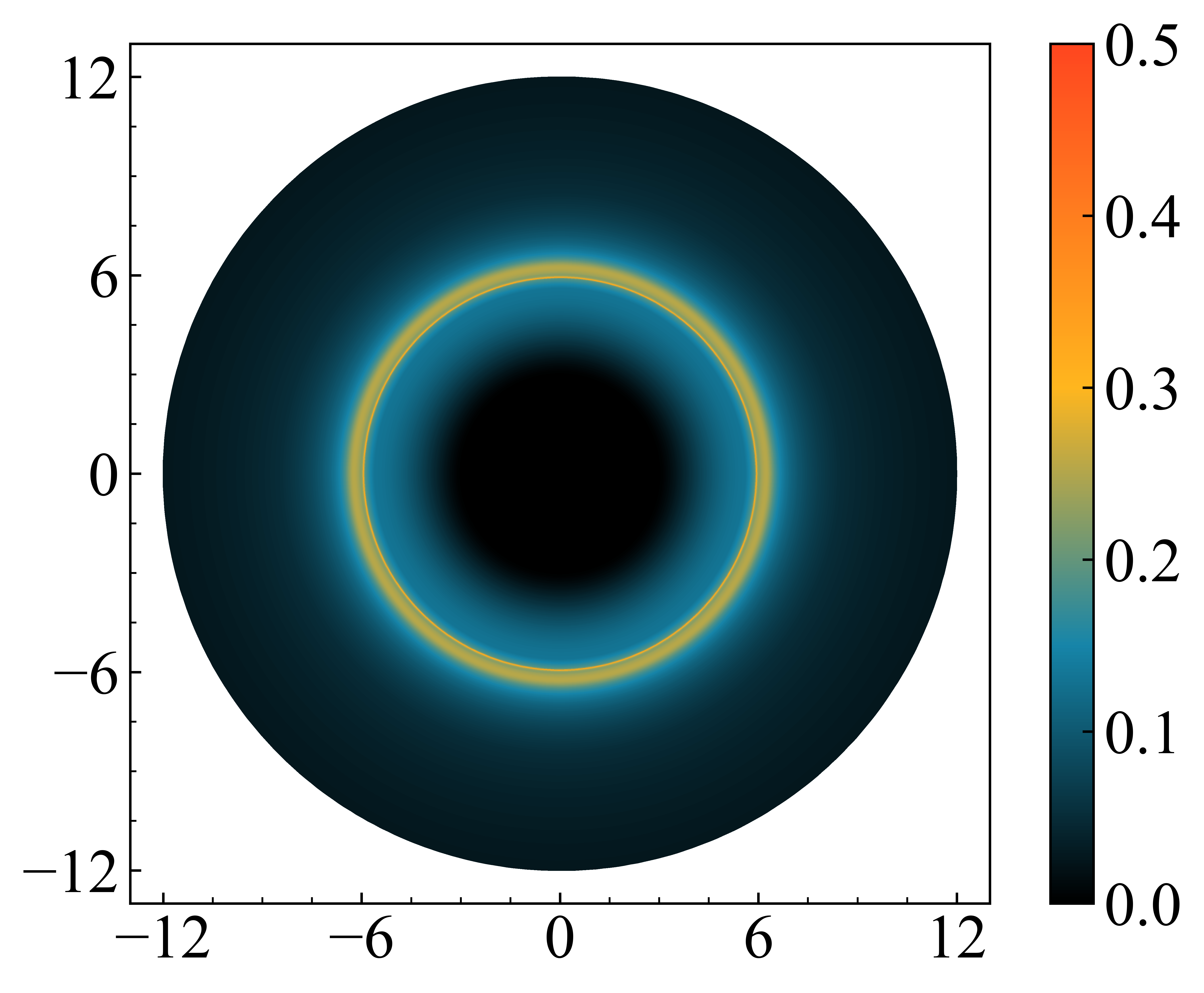}
}
\caption{Observational appearances of the bumblebee black hole for the moderate decay function.}
\label{fig:Iemi_mod}
\end{figure}

The observational outcomes for this model are presented in Fig.~\ref{fig:Iemi_mod}. As shown in the left column, the intrinsic emission profile extends smoothly to the horizon without a sharp cutoff. The corresponding observed intensity (middle column) exhibits a broad and complex structure. Unlike the previous models where distinct peaks could be identified, here the direct emission, lensing ring, and photon ring are merged into a single, continuous luminosity distribution. The intensity profile rises gradually from the outer region, culminating in a prominent peak near the critical curve before dropping sharply at the shadow boundary.

A quantitative analysis of the intensity profiles reveals a significant dependence on the Lorentz violation parameter $\chi$. For $\chi=0.1$, the peak observed intensity is approximately $0.47$, located at $b \approx 5.46M$. In the case of $\chi=0.3$, the peak intensity drops to approximately $0.34$, and the position of the maximum shifts outward to $b \approx 5.95M$. This reduction in peak brightness is more pronounced than in the other models, suggesting that the near-horizon radiation is strongly affected by the geometric dilation induced by $\chi$. The images in the right column further illustrate this effect: the observable ring in the $\chi=0.3$ case appears significantly wider and dimmer compared to the $\chi=0.1$ case. This confirms that for accretion disks extending to the horizon, a non--zero bumblebee parameter $\chi$ would manifest observationally as a larger, more diffuse, and less luminous shadow boundary.

The numerical results from the above three models \eqref{model_2nd}-\eqref{model_mod} consistently exhibit two primary observational trends associated with the increase of the Lorentz violation parameter $\chi$: an outward shift of the optical rings and a simultaneous decrease in the observed peak intensity. These features can be directly understood by analyzing how the parameter $\chi$ modifies the spacetime geometry and the photon propagation equations.

The expansion of the ring structures in the impact parameter space is a direct consequence of the modified critical curve. As derived in Eq.~\eqref{Veff}, the effective potential barrier is modulated by the factor $(1+\chi)^{-1}$. This modification scales the critical impact parameter as $b_p \propto \sqrt{1+\chi}$. Consequently, for a larger $\chi$, the threshold for photon capture increases, causing the critical trajectories (and thus the photon and lensing rings) to map onto larger radial coordinates on the observer's screen. This explains why the boundary of the shadow appears larger in the $\chi=0.3$ case compared to the $\chi=0.1$ case, regardless of the emission profile structure.

On the other hand, the reduction in the observed intensity can be attributed to the dependence of the gravitational redshift on the metric coefficients. The observed specific intensity is related to the intrinsic emission via the redshift factor $g = \sqrt{-g_{tt}}$. It is evident that the redshift factor scales inversely with the square root of $1+\chi$. Therefore, an increase in $\chi$ leads to a smaller value of $g$ for a fixed radial emitter. Since the total observed intensity scales as $I_{obs} \propto g^4$, this dependency results in a significant reduction in the received flux.

Therefore, the non-zero vacuum expectation value of the bumblebee field imposes a dual modulation on the black hole image: it scales the apparent size of the shadow and attenuates the brightness of the surrounding accretion flow. In this manner, these correlated variations provide a specific observational signature for the violation of Lorentz symmetry in the strong gravity regime.


\section{Shadows and photon spheres with spherical accretions}

In addition to the thin disk scenario, astrophysical black holes are often surrounded by geometrically thick, optically thin matter distributions, such as spherical accretion flows. Unlike the discrete ring structures formed by thin disks, spherical accretions produce a continuous luminosity profile, where the shadow appears as a central brightness depression. In this section, we extend our analysis to the bumblebee black hole surrounded by spherical accretions, considering both static and radially infalling matter configurations.

\subsection{Static spherical accretion}

\begin{figure}[t]
\centering
\includegraphics[width=0.48\textwidth]{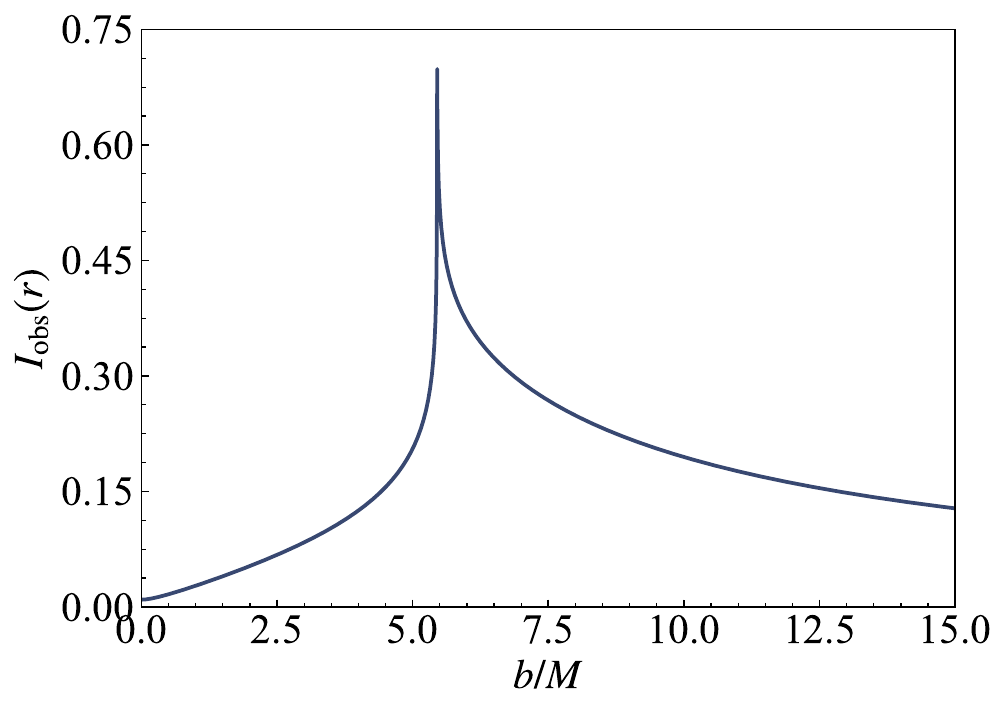}
\hfill
\includegraphics[width=0.48\textwidth]{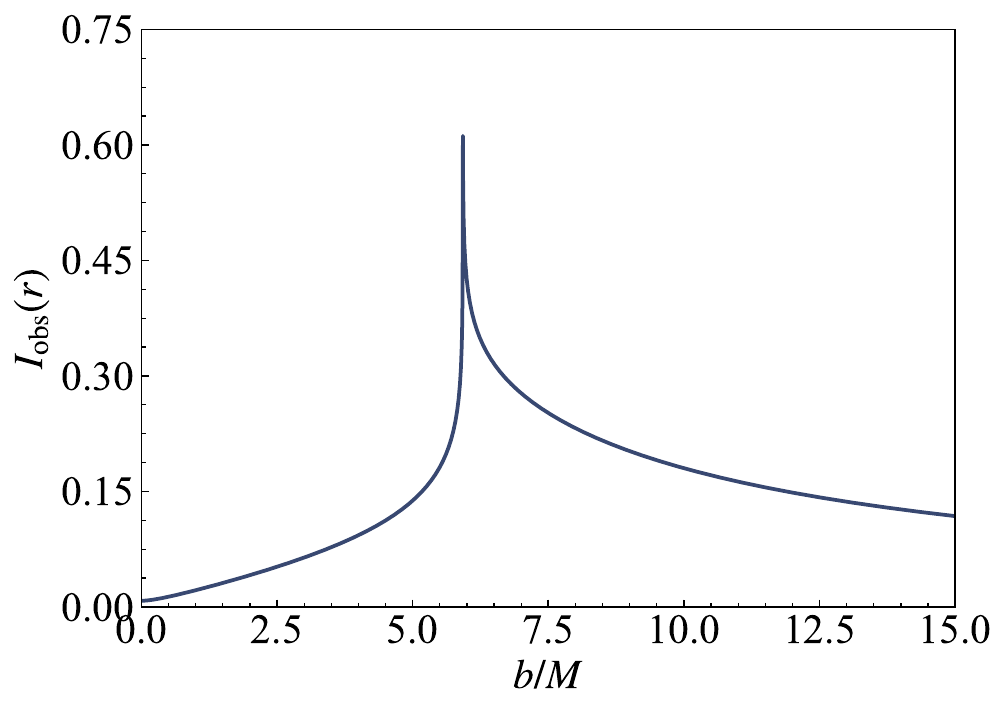}
\\
\subfloat[$\chi=0.1$\label{fig:traj2_1}]{
\includegraphics[width=0.48\textwidth]{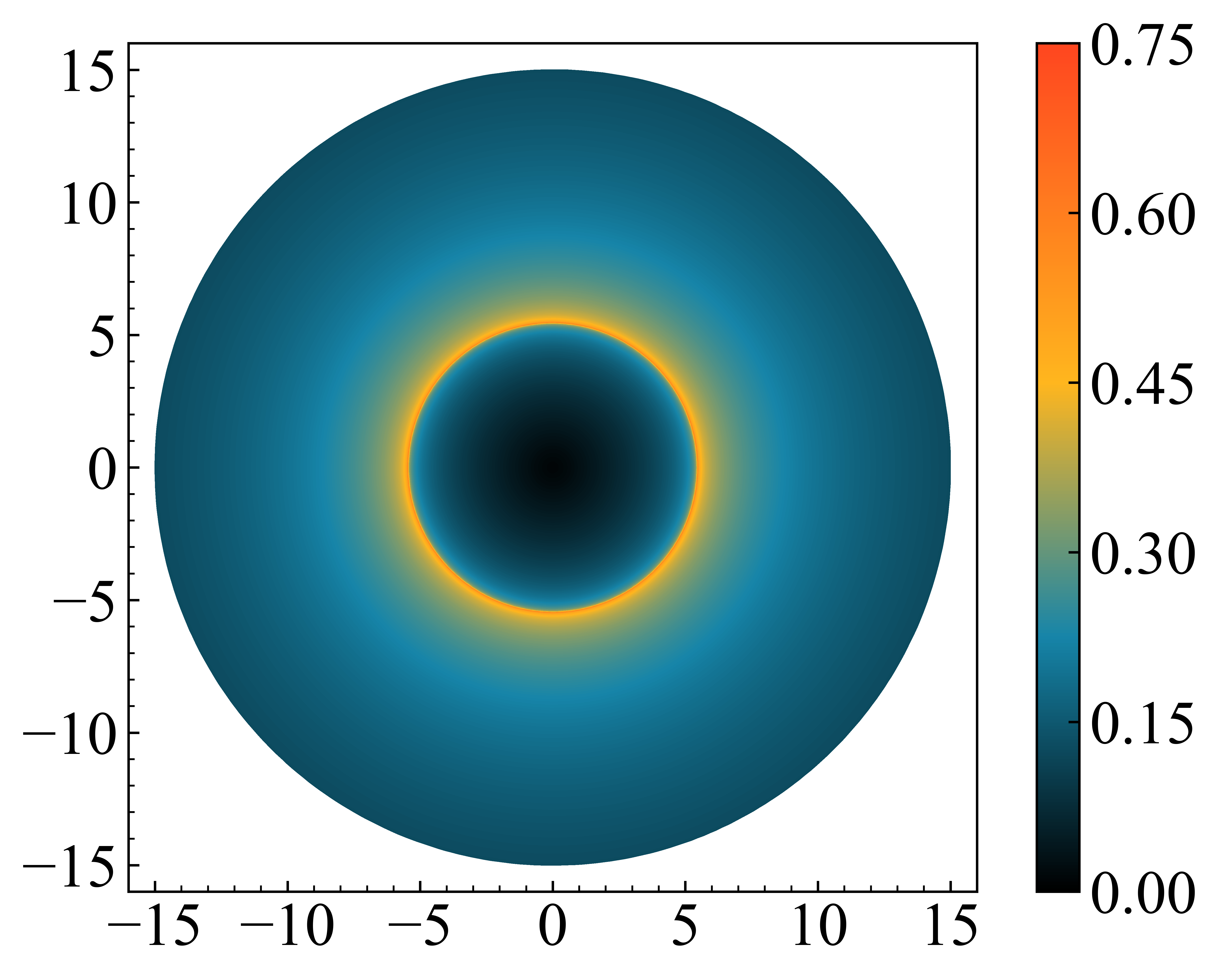}
}
\hfill
\subfloat[$\chi=0.3$\label{fig:traj2_2}]{
\includegraphics[width=0.48\textwidth]{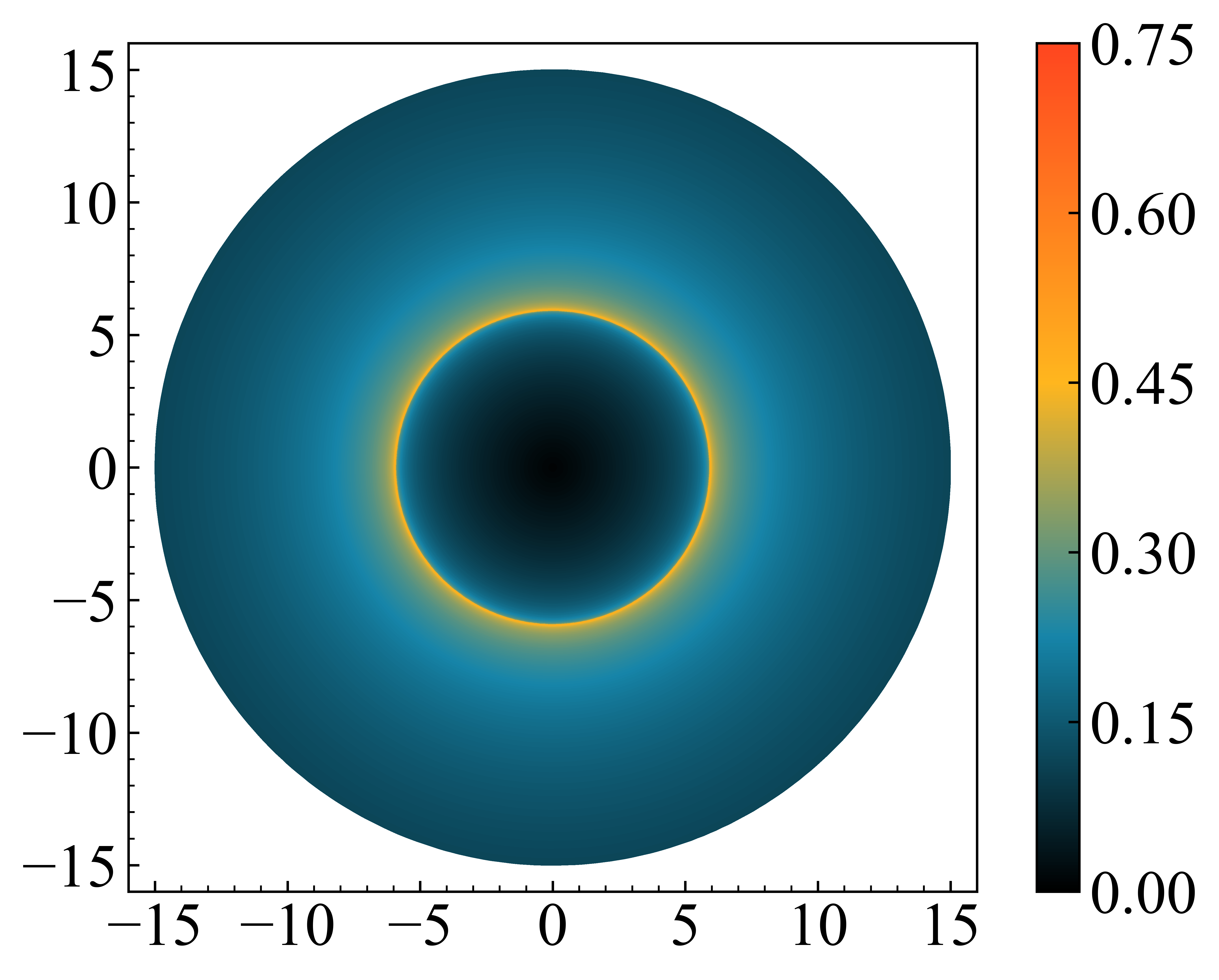}
}
\caption{Observational appearances of the bumblebee black hole for the static spherical accretion.}
\label{fig:Iemi_static}
\end{figure}

We first consider a static, optically thin accretion surrounding the black hole. For an observer at infinity, the observed specific intensity $I_{obs}(\nu_{obs})$ at a specific frequency $\nu_{obs}$ is given by the line-of-sight integral of the emissivity \cite{jaroszynski1997optics,bambi2013can,fathi2023observational}:
\begin{align}
I_{obs}(\nu_{obs}) = \int_{\gamma} g^3 j(\nu_{emi}) \mathrm{d}l_{\text{prop}},
\end{align}
where $g = \nu_{obs}/\nu_{emi}$ is the redshift factor, $j(\nu_{emi})$ is the emissivity per unit volume in the emitter's rest frame, and $\mathrm{d}l_{\text{prop}}$ is the infinitesimal proper length along the photon path $\gamma$. Assuming a monochromatic emission with a radial inverse-square decay profile, the emissivity is defined as 
\begin{align}
j(\nu_{emi}) \propto \dfrac{\delta(\nu_{emi} - \nu_{*})}{r^2}.
\end{align}

For a static observer at infinity and a static matter distribution, the four-velocity of the emitter is 
\begin{align}
u^{\mu}_{emi} = \left(\dfrac{1}{\sqrt{-g_{tt}}}, 0, 0, 0\right).
\end{align}
Consequently, the redshift factor is determined purely by
\begin{align}
g = \dfrac{1}{\sqrt{1+\chi}}\sqrt{1 - \dfrac{2M}{r}}.
\end{align}
The proper length element in the new bumblebee spacetime can be derived from the new bumblebee metric Eq.~\ref{metric} as:
\begin{align}
\mathrm{d}l_{\text{prop}} = \sqrt{\dfrac{1+\chi}{1-\frac{2M}{r}} + r^2 \left(\dfrac{\mathrm{d}\phi}{\mathrm{d}r}\right)^2} \mathrm{d}r.
\end{align}
By substituting the orbital equation $\mathrm{d}\phi/\mathrm{d}r$ derived in Eq.~\eqref{eq:orbit_u}, we numerically integrate the observed intensity as a function of the impact parameter $b$.

The resulting observational signatures are presented in Fig.~\ref{fig:Iemi_static}. The top row displays the intensity profiles $I_{obs}(b)$, which rise smoothly to a sharp peak at the critical impact parameter $b_p$ before decaying at larger radii. This peak corresponds to the photon sphere, where the optical path length through the emitting medium is maximized due to the strong gravitational lensing. Inside the critical curve ($b < b_p$), the intensity drops rapidly, forming the central shadow.

A comparison between the $\chi=0.1$ (left column) and $\chi=0.3$ (right column) cases reveals the dual impact of the Lorentz violation parameter. First, the position of the maximum brightness shifts outward from $b \approx 5.45M$ to $b \approx 5.92M$, confirming the enlargement of the shadow size. Second, the magnitude of the peak intensity decreases noticeably, dropping from approximately $0.70$ to $0.61$. This suppression is a direct consequence of the modified redshift factor $g \propto (1+\chi)^{-1/2}$, which uniformly dims the observed radiation. The images in the bottom row visualize these effects: the shadow in the $\chi=0.3$ case is larger but surrounded by a less luminous photon ring compared to the $\chi=0.1$ case.


\subsection{Infalling spherical accretion}

\begin{figure}[t]
\centering
\includegraphics[width=0.48\textwidth]{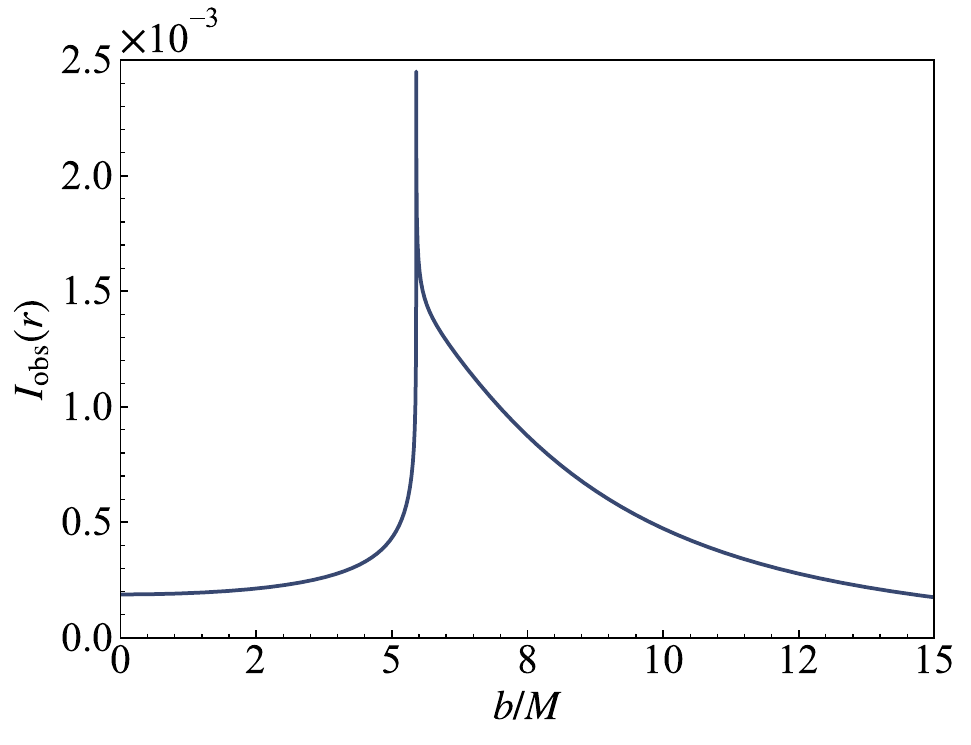}
\hfill
\includegraphics[width=0.48\textwidth]{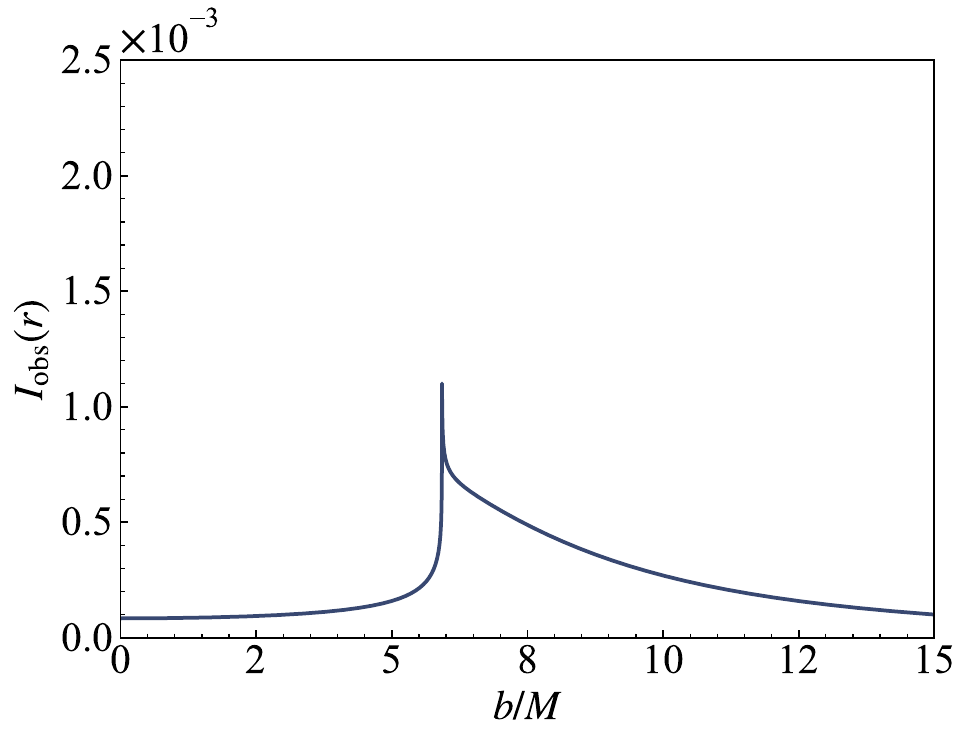}
\\
\subfloat[$\chi=0.1$\label{fig:traj2_1}]{
\includegraphics[width=0.48\textwidth]{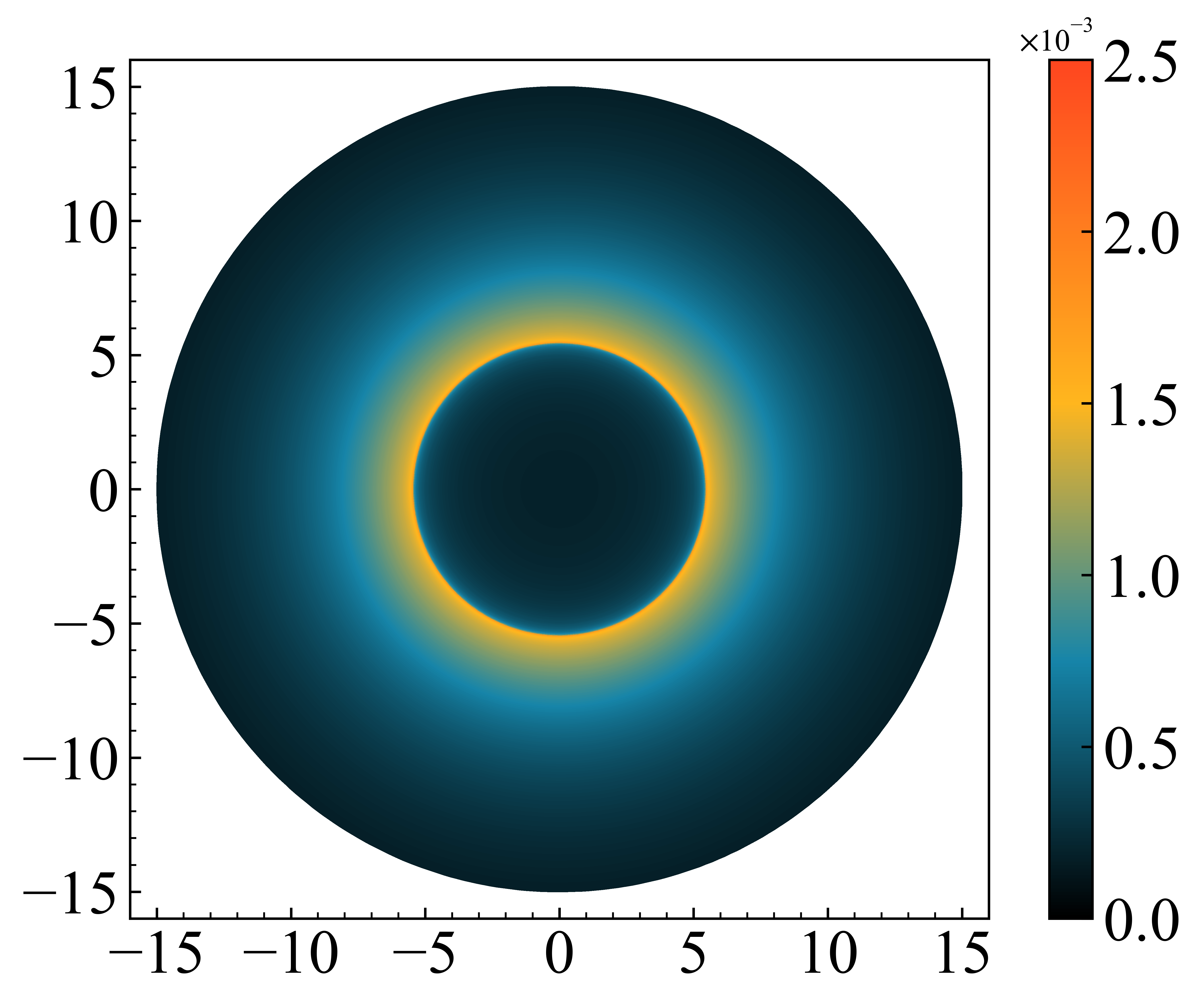}
}
\hfill
\subfloat[$\chi=0.3$\label{fig:traj2_2}]{
\includegraphics[width=0.48\textwidth]{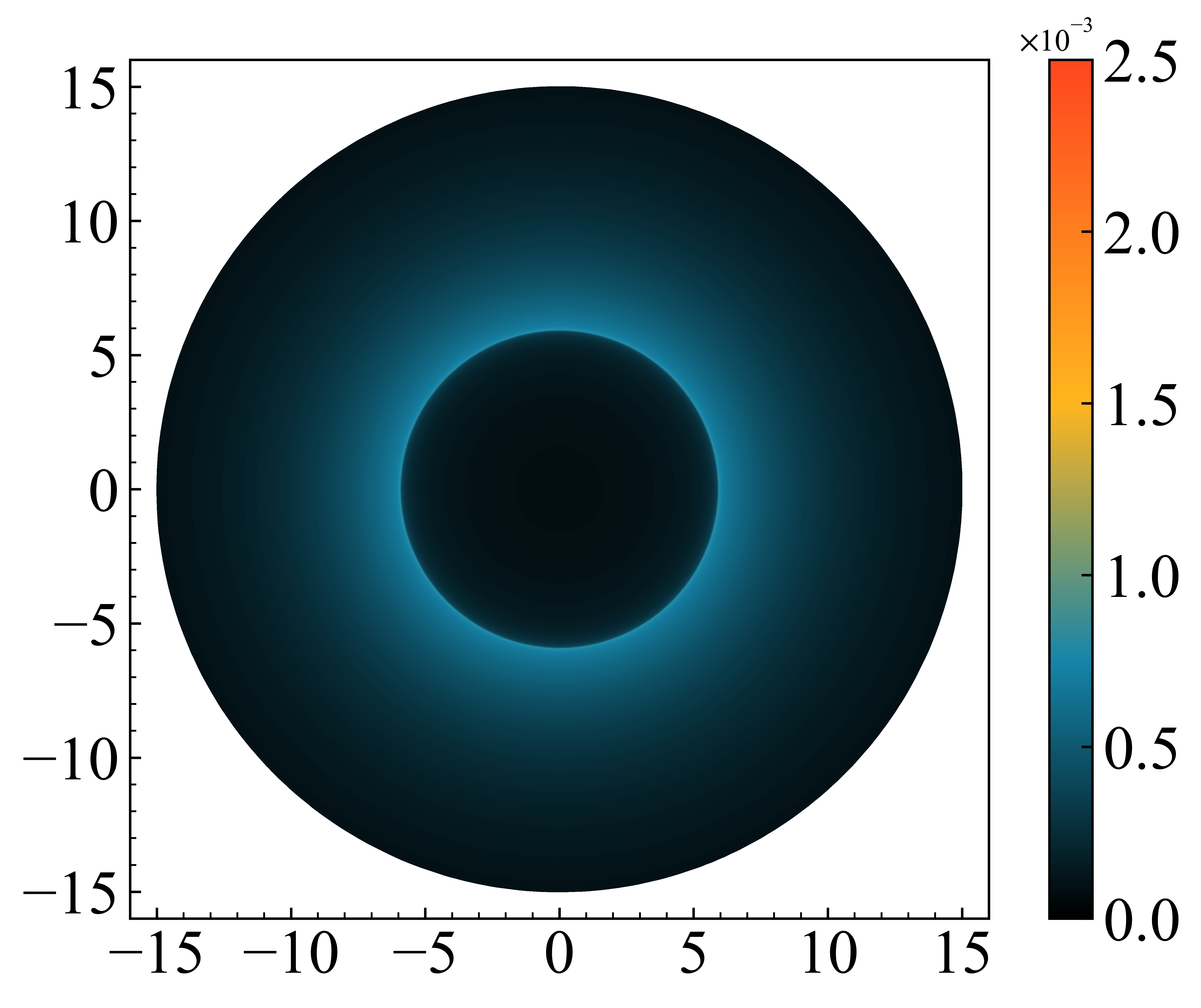}
}
\caption{Observational appearances of the bumblebee black hole for the infalling spherical accretion.}
\label{fig:Iemi_infalling}
\end{figure}

A more realistic scenario involves matter dynamically accreting onto the black hole. We consider a radial free--fall model where the gas starts from rest at infinity. The four--velocity of the infalling fluid $u^{\mu}_{emi}$ is given by \cite{li2021observational,li2021shadows,fathi2023observational,bambi2013can}:
\begin{align}
u^{t}_{emi} = \dfrac{1}{-g_{tt}}, \quad u^{r}_{emi} = -\sqrt{1 - g_{tt}}.
\end{align}
In this dynamical context, the redshift factor $g$ must account for both gravitational redshift and the Doppler effect arising from the bulk motion of the fluid. It is expressed as \cite{bambi2013can}:
\begin{align}
g = \dfrac{k_{\mu} u^{\mu}_{obs}}{k_{\nu} u^{\nu}_{emi}},
\end{align}
where $k^{\mu} = \dot{x}^{\mu}$ is the photon four--momentum and $u^{\mu}_{obs} = (1, 0, 0, 0)$ corresponds to the static observer at infinity.

The observational results for the infalling accretion are presented in Fig.~\ref{fig:Iemi_infalling}. Similar to the static case, the observed intensity profiles (top row) exhibit a sharp peak at the critical impact parameter. However, the overall luminosity is orders of magnitude lower due to the Doppler effect, as the emitting matter moves away from the observer at high velocities near the horizon. 

Quantitatively, for the case of $\chi=0.1$, the maximum intensity reaches approximately $0.00245$ at $b \approx 5.45M$. When the Lorentz violation parameter increases to $\chi=0.3$, the peak intensity drops further to $1.1\times10^{-3}$, with the peak position shifting to $b \approx 5.93M$. This significant reduction confirms that the combination of the new bumblebee geometry and the kinematic redshift strongly suppresses the observed flux. Furthermore, inside the critical curve ($b < b_p$), the intensity decays much faster than in the static case, rendering the central shadow significantly darker. This feature is clearly visible in Fig.~\ref{fig:Iemi_infalling} (bottom row), where the contrast between the photon ring and the central dark region is enhanced compared to the static scenario.


\section{Conclusion}\label{Sec:Conclusion}

This work examined how the new static bumblebee black hole modified photon dynamics and the observed emission from different accretion models. The Lorentz--violating deformation shifted the critical impact parameter outward, which enlarged the shadow and changed the positions of the direct emission, lensing rings, and photon rings. Ray--tracing simulations confirmed that all ring structures moved to higher impact parameters as the parameter increased, and the separations between the rings widened accordingly. In the thin--disk analysis, the outward shift of the critical curve altered the mapping from the emitting radii to the observer’s screen, producing broader arcs and modifying their relative brightness.

For static spherical accretion, the luminosity profiles showed a pronounced central dimming, with the depression becoming wider for larger values of the Lorentz--violating parameter. In the infalling configuration, the combined gravitational and Doppler redshifts suppressed the flux by several orders of magnitude. The peak intensity decreased noticeably when the parameter was raised, and its location moved outward along the impact--parameter axis. Inside the critical curve, the intensity decayed more rapidly than in the static case, which rendered the central region significantly darker and increased the contrast between the photon ring and the interior shadow. The numerical examples illustrated this trend clearly, such as the reduction of the peak intensity for the infalling model when the parameter increased from 0.1 to 0.3, together with the shift in the peak position.


\section*{Acknowledgments}
\hspace{0.5cm} A.A.A.F. is supported by Conselho Nacional de Desenvolvimento Cient\'{\i}fico e Tecnol\'{o}gico (CNPq) and Fundação de Apoio à Pesquisa do Estado da Paraíba (FAPESQ), project numbers 150223/2025-0 and 1951/2025. The authors also thank I.~P.~Lobo for the comments and suggestions provided while this manuscript was being prepared.

\section*{Data Availability Statement}

Data Availability Statement: No Data associated with the manuscript

\bibliographystyle{ieeetr}
\bibliography{main}

\begin{thebibliography}{100}

\bibitem{colladay1997cpt}
D.~Colladay and V.~A. Kosteleck{\`y}, ``Cpt violation and the standard model,''
  {\em Physical Review D}, vol.~55, no.~11, p.~6760, 1997.

\bibitem{kostelecky1989spontaneous}
V.~A. Kosteleck{\`y} and S.~Samuel, ``Spontaneous breaking of lorentz symmetry
  in string theory,'' {\em Physical Review D}, vol.~39, no.~2, p.~683, 1989.

\bibitem{kostelecky2004gravity}
V.~A. Kosteleck{\`y}, ``Gravity, lorentz violation, and the standard model,''
  {\em Physical Review D}, vol.~69, no.~10, p.~105009, 2004.

\bibitem{kostelecky2011data}
V.~A. Kosteleck{\`y} and N.~Russell, ``Data tables for lorentz and cpt
  violation,'' {\em Reviews of Modern Physics}, vol.~83, no.~1, pp.~11--31,
  2011.

\bibitem{kostelecky1999constraints}
V.~A. Kosteleck{\`y} and C.~D. Lane, ``Constraints on lorentz violation from
  clock-comparison experiments,'' {\em Physical Review D}, vol.~60, no.~11,
  p.~116010, 1999.

\bibitem{bluhm2005spontaneous}
R.~Bluhm and V.~A. Kosteleck{\`y}, ``Spontaneous lorentz violation,
  nambu-goldstone modes, and gravity,'' {\em Physical Review D—Particles,
  Fields, Gravitation, and Cosmology}, vol.~71, no.~6, p.~065008, 2005.

\bibitem{Bluhm:2023kph}
R.~Bluhm and Y.~Zhi, ``{Spontaneous and Explicit Spacetime Symmetry Breaking in
  Einstein{\textendash}Cartan Theory with Background Fields},'' {\em Symmetry},
  vol.~16, no.~1, p.~25, 2024.

\bibitem{Bluhm:2019ato}
R.~Bluhm, H.~Bossi, and Y.~Wen, ``{Gravity with explicit spacetime symmetry
  breaking and the Standard-Model Extension},'' {\em Phys. Rev. D}, vol.~100,
  no.~8, p.~084022, 2019.

\bibitem{Maluf:2014dpa}
R.~V. Maluf, C.~A.~S. Almeida, R.~Casana, and M.~M. Ferreira, Jr.,
  ``{Einstein-Hilbert graviton modes modified by the Lorentz-violating
  bumblebee Field},'' {\em Phys. Rev. D}, vol.~90, no.~2, p.~025007, 2014.

\bibitem{Maluf:2013nva}
R.~V. Maluf, V.~Santos, W.~T. Cruz, and C.~A.~S. Almeida, ``{Matter-gravity
  scattering in the presence of spontaneous Lorentz violation},'' {\em Phys.
  Rev. D}, vol.~88, no.~2, p.~025005, 2013.

\bibitem{bluhm2008spontaneous}
R.~Bluhm, S.-H. Fung, and V.~A. Kosteleck{\`y}, ``Spontaneous lorentz and
  diffeomorphism violation, massive modes, and gravity,'' {\em Physical Review
  D—Particles, Fields, Gravitation, and Cosmology}, vol.~77, no.~6,
  p.~065020, 2008.

\bibitem{kostelecky1991photon}
V.~A. Kosteleck{\`y} and S.~Samuel, ``Photon and graviton masses in string
  theories,'' {\em Physical Review Letters}, vol.~66, no.~14, p.~1811, 1991.

\bibitem{jacobson2004einstein}
T.~Jacobson and D.~Mattingly, ``Einstein-aether waves,'' {\em Physical Review
  D}, vol.~70, no.~2, p.~024003, 2004.

\bibitem{Liu:2022dcn}
W.~Liu, X.~Fang, J.~Jing, and J.~Wang, ``{QNMs of slowly rotating
  Einstein{\textendash}Bumblebee black hole},'' {\em Eur. Phys. J. C}, vol.~83,
  no.~1, p.~83, 2023.

\bibitem{Bertolami:2005bh}
O.~Bertolami and J.~Paramos, ``{The Flight of the bumblebee: Vacuum solutions
  of a gravity model with vector-induced spontaneous Lorentz symmetry
  breaking},'' {\em Phys. Rev. D}, vol.~72, p.~044001, 2005.

\bibitem{Casana:2017jkc}
R.~Casana, A.~Cavalcante, F.~P. Poulis, and E.~B. Santos, ``{Exact
  Schwarzschild-like solution in a bumblebee gravity model},'' {\em Phys. Rev.
  D}, vol.~97, no.~10, p.~104001, 2018.

\bibitem{Liu:2024wpa}
W.~Liu, C.~Wen, and J.~Wang, ``{Lorentz violation alleviates gravitationally
  induced entanglement degradation},'' {\em JHEP}, vol.~01, p.~184, 2025.

\bibitem{AraujoFilho:2025hkm}
A.~A. Ara{\'u}jo~Filho, ``{How does non-metricity affect particle creation and
  evaporation in bumblebee gravity?},'' {\em JCAP}, vol.~06, p.~026, 2025.

\bibitem{AraujoFilho:2024ctw}
A.~A. Ara{\'u}jo~Filho, ``{Particle creation and evaporation in Kalb-Ramond
  gravity},'' {\em JCAP}, vol.~04, p.~076, 2025.

\bibitem{Neves:2022qyb}
J.~C.~S. Neves, ``{Kasner cosmology in bumblebee gravity},'' {\em Annals
  Phys.}, vol.~454, p.~169338, 2023.

\bibitem{Neves:2024ggn}
J.~C.~S. Neves and F.~G. Gardim, ``{Stars and quark stars in bumblebee
  gravity},'' {\em Annals Phys.}, vol.~475, p.~169950, 2025.

\bibitem{Liang:2022hxd}
D.~Liang, R.~Xu, X.~Lu, and L.~Shao, ``{Polarizations of gravitational waves in
  the bumblebee gravity model},'' {\em Phys. Rev. D}, vol.~106, no.~12,
  p.~124019, 2022.

\bibitem{amarilo2024gravitational}
K.~M. Amarilo, M.~B. Ferreira~Filho, A.~A. Ara{\'u}jo~Filho, and J.~A. A.~S.
  Reis, ``Gravitational waves effects in a lorentz--violating scenario,'' {\em
  Physics Letters B}, vol.~855, p.~138785, 2024.

\bibitem{Maluf:2020kgf}
R.~V. Maluf and J.~C.~S. Neves, ``{Black holes with a cosmological constant in
  bumblebee gravity},'' {\em Phys. Rev. D}, vol.~103, no.~4, p.~044002, 2021.

\bibitem{Filho:2022yrk}
A.~A.~A. Filho, J.~R. Nascimento, A.~Y. Petrov, and P.~J. Porf{\'\i}rio,
  ``{Vacuum solution within a metric-affine bumblebee gravity},'' {\em Phys.
  Rev. D}, vol.~108, no.~8, p.~085010, 2023.

\bibitem{AraujoFilho:2024ykw}
A.~A. Ara{\'u}jo~Filho, J.~R. Nascimento, A.~Y. Petrov, and P.~J.
  Porf{\'\i}rio, ``{An exact stationary axisymmetric vacuum solution within a
  metric-affine bumblebee gravity},'' {\em JCAP}, vol.~07, p.~004, 2024.

\bibitem{AraujoFilho:2025rvn}
A.~A. Ara{\'u}jo~Filho, N.~Heidari, I.~P. Lobo, Y.~Shi, and F.~S.~N. Lobo,
  ``{The Flight of the Bumblebee in a Non-Commutative Geometry: A New Black
  Hole Solution},'' 9 2025.

\bibitem{AraujoFilho:2025jcu}
A.~A. Ara{\'u}jo~Filho, N.~Heidari, and I.~P. Lobo, ``{A non-commutative
  Kalb-Ramond black hole},'' {\em JCAP}, vol.~09, p.~076, 2025.

\bibitem{Ovgun:2018xys}
A.~{\"O}vg{\"u}n, K.~Jusufi, and {\.I}.~Sakall{\i}, ``{Exact traversable
  wormhole solution in bumblebee gravity},'' {\em Phys. Rev. D}, vol.~99,
  no.~2, p.~024042, 2019.

\bibitem{Magalhaes:2025nql}
R.~B. Magalh{\~a}es, L.~A. Lessa, and M.~M. Ferreira, ``{Wormholes in
  Lorentz-violating gravity},'' 5 2025.

\bibitem{Magalhaes:2025lti}
R.~B. Magalh{\~a}es, L.~A. Lessa, and R.~Casana, ``{Lorentz-violating
  wormholes: The role of the matter coupled to Lorentz-violating fields},'' 7
  2025.

\bibitem{AraujoFilho:2024iox}
A.~A. Ara{\'u}jo~Filho, J.~A. A.~S. Reis, and A.~{\"O}vg{\"u}n, ``{Modified
  particle dynamics and thermodynamics in a traversable wormhole in bumblebee
  gravity},'' {\em Eur. Phys. J. C}, vol.~85, no.~1, p.~83, 2025.

\bibitem{Pereira:2025xnw}
C.~F.~S. Pereira, M.~V. d.~S. Silva, H.~Belich, D.~C.~Rodrigues, J.~C. Fabris,
  and M.~E. Rodrigues, ``{Black-bounce solutions in a k-essence theory under
  the effects of bumblebee gravity},'' {\em Phys. Rev. D}, vol.~111, no.~12,
  p.~124005, 2025.

\bibitem{Shi:2025plr}
Y.~Shi and A.~A. Ara{\'u}jo~Filho, ``Effects of bumblebee gravity on neutrino
  motion,'' {\em Journal of Cosmology and Astroparticle Physics}, vol.~2025,
  no.~11, p.~045, 2025.

\bibitem{Shi:2025ywa}
Y.~Shi and A.~A. Ara{\'u}jo~Filho, ``The role of non-metricity on neutrino
  behavior in bumblebee gravity,'' {\em arXiv preprint arXiv:2505.12551}, 2025.

\bibitem{Shi:2025rfq}
Y.~Shi and A.~A. Ara{\'u}jo~Filho, ``{Influence of a Kalb-Ramond black hole on
  neutrino behavior},'' {\em JHEP}, vol.~08, p.~028, 2025.

\bibitem{Liu:2025oho}
J.-Z. Liu, S.-P. Wu, S.-W. Wei, and Y.-X. Liu, ``{Exact Black Hole Solutions in
  Bumblebee Gravity with Lightlike or Spacelike VEVS},'' 10 2025.

\bibitem{Zhu:2025fiy}
J.~Zhu and H.~Li, ``{Full Classification of Static Spherical Vacuum Solutions
  to Bumblebee Gravity with General VEVs},'' 11 2025.

\bibitem{AraujoFilho:2025zaj}
A.~A. Ara{\'u}jo~Filho, N.~Heidari, I.~P. Lobo, and V.~B. Bezerra,
  ``{Gravitational aspects of a new bumblebee black hole},'' 11 2025.

\bibitem{Shi:2025tvu}
Y.~Shi and A.~A. Ara{\'u}jo~Filho, ``{Neutrino oscillations induced by a new
  bumblebee black hole},'' 11 2025.

\bibitem{Kumar:2025bim}
A.~Kumar, S.~U. Islam, and S.~G. Ghosh, ``{Probing Lorentz Symmetry Violation
  through Lensing Observables of Rotating Black Holes},'' 8 2025.

\bibitem{Cunningham}
C.~T. Cunningham and J.~M. Bardeen, ``{The Optical Appearance of a Star
  Orbiting an Extreme Kerr Black Hole},'' {\em The Astrophysical Journal},
  vol.~183, pp.~237--264, 1973.

\bibitem{Falcke:1999pj}
H.~Falcke, F.~Melia, and E.~Agol, ``Viewing the shadow of the black hole at the
  galacticcenter,'' {\em The Astrophysical Journal}, vol.~528, no.~1, p.~L13,
  1999.

\bibitem{Afrin:2024khy}
M.~Afrin, S.~G. Ghosh, and A.~Wang, ``{Testing EGB gravity coupled to bumblebee
  field and black hole parameter estimation with EHT observations},'' {\em
  Physics of the Dark Universe}, vol.~46, p.~101642, 2024.

\bibitem{Khodadi:2024ubi}
M.~Khodadi, S.~Vagnozzi, and J.~T. Firouzjaee, ``{Event Horizon Telescope
  observations exclude compact objects in baseline mimetic gravity},'' {\em
  Scientific Reports}, vol.~14, no.~1, p.~26932, 2024.

\bibitem{Allahyari:2019jqz}
A.~Allahyari, M.~Khodadi, S.~Vagnozzi, and D.~F. Mota, ``{Magnetically charged
  black holes from non-linear electrodynamics and the Event Horizon
  Telescope},'' {\em Journal of Cosmology and Astroparticle Physics,}, vol.~02,
  p.~003, 2020.

\bibitem{Afrin:2021wlj}
M.~Afrin and S.~G. Ghosh, ``Testing horndeski gravity from eht observational
  results for rotating black holes,'' {\em The Astrophysical Journal},
  vol.~932, no.~1, p.~51, 2022.

\bibitem{Nojiri:2024txy}
S.~Nojiri and S.~D. Odintsov, ``{Improving mimetic gravity with non-trivial
  scalar potential: Cosmology, black holes, shadow and photon sphere},'' {\em
  Physics of the Dark Universe}, vol.~46, p.~101669, 2024.

\bibitem{Afrin:2021imp}
M.~Afrin, R.~Kumar, and S.~G. Ghosh, ``Parameter estimation of hairy kerr black
  holes from its shadow and constraints from {M87$^*$},'' {\em Monthly Notices
  of the Royal Astronomical Society}, vol.~504, no.~4, pp.~5927--5940, 2021.

\bibitem{Nojiri:2024qgx}
S.~Nojiri and S.~D. Odintsov, ``{Black holes and their shadows in {$F(R)$}
  gravity},'' {\em Physics of the Dark Universe}, vol.~47, p.~101785, 2025.

\bibitem{Bambi:2019tjh}
C.~Bambi, K.~Freese, S.~Vagnozzi, and L.~Visinelli, ``{Testing the rotational
  nature of the supermassive object {M87$^*$} from the circularity and size of
  its first image},'' {\em Physical Review D}, vol.~100, no.~4, p.~044057,
  2019.

\bibitem{Khodadi:2021gbc}
M.~Khodadi, G.~Lambiase, and D.~F. Mota, ``{No-hair theorem in the wake of
  Event Horizon Telescope},'' {\em Journal of Cosmology and Astroparticle
  Physics,}, vol.~09, p.~028, 2021.

\bibitem{Liu:2024lve}
W.~Liu, D.~Wu, and J.~Wang, ``{Shadow of slowly rotating Kalb-Ramond black
  holes},'' {\em Journal of Cosmology and Astroparticle Physics,}, vol.~05,
  p.~017, 2025.

\bibitem{Khodadi:2022pqh}
M.~Khodadi and G.~Lambiase, ``{Probing Lorentz symmetry violation using the
  first image of Sagittarius A*: Constraints on standard-model extension
  coefficients},'' {\em Physical Review D}, vol.~106, no.~10, p.~104050, 2022.

\bibitem{Kumar:2020hgm}
R.~Kumar, S.~G. Ghosh, and A.~Wang, ``{Gravitational deflection of light and
  shadow cast by rotating Kalb-Ramond black holes},'' {\em Physical Review D},
  vol.~101, no.~10, p.~104001, 2020.

\bibitem{Vagnozzi:2019apd}
S.~Vagnozzi and L.~Visinelli, ``{Hunting for extra dimensions in the shadow of
  {M87$^*$}},'' {\em Physical Review D}, vol.~100, no.~2, p.~024020, 2019.

\bibitem{Nojiri:2024nlx}
S.~Nojiri and S.~D. Odintsov, ``{Black holes, photon sphere, and cosmology in
  ghost-free {$f(G)$} gravity},'' {\em Physics of the Dark Universe}, vol.~46,
  p.~101702, 2024.

\bibitem{Fu:2021fxn}
Q.-M. Fu and X.~Zhang, ``Gravitational lensing by a black hole in effective
  loop quantum gravity,'' {\em Physical Review D}, vol.~105, no.~6, p.~064020,
  2022.

\bibitem{Liu:2024soc}
W.~Liu, D.~Wu, and J.~Wang, ``{Light rings and shadows of static black holes in
  effective quantum gravity},'' {\em Phys. Lett. B}, vol.~858, p.~139052, 2024.

\bibitem{40}
S.~E. Gralla, D.~E. Holz, and R.~M. Wald, ``{Black Hole Shadows, Photon Rings,
  and Lensing Rings},'' {\em Phys. Rev. D}, vol.~100, no.~2, p.~024018, 2019.

\bibitem{41}
N.~Tsukamoto, Z.~Li, and C.~Bambi, ``{Constraining the spin and the deformation
  parameters from the black hole shadow},'' {\em JCAP}, vol.~06, p.~043, 2014.

\bibitem{42}
N.~Tsukamoto, ``{Black hole shadow in an asymptotically-flat, stationary, and
  axisymmetric spacetime: The Kerr-Newman and rotating regular black holes},''
  {\em Phys. Rev. D}, vol.~97, no.~6, p.~064021, 2018.

\bibitem{Gogoi:2024eyw}
D.~J. Gogoi, J.~Bora, F.~Studni{\v{c}}ka, and H.~Hassanabadi, ``{Optical,
  dynamic and topological characteristics of deformed schwarzschild black
  holes},'' {\em JCAP}, vol.~08, p.~009, 2025.

\bibitem{Zare:2024dtf}
S.~Zare, L.~M. Nieto, X.-H. Feng, S.-H. Dong, and H.~Hassanabadi, ``{Shadows,
  rings and optical appearance of a magnetically charged regular black hole
  illuminated by various accretion disks},'' {\em JCAP}, vol.~08, p.~041, 2024.

\bibitem{43}
X.-X. Zeng and H.-Q. Zhang, ``{Influence of quintessence dark energy on the
  shadow of black hole},'' {\em Eur. Phys. J. C}, vol.~80, no.~11, p.~1058,
  2020.

\bibitem{44}
X.-X. Zeng, H.-Q. Zhang, and H.~Zhang, ``{Shadows and photon spheres with
  spherical accretions in the four-dimensional Gauss{\textendash}Bonnet black
  hole},'' {\em Eur. Phys. J. C}, vol.~80, no.~9, p.~872, 2020.

\bibitem{46}
K.-J. He, S.~Guo, S.-C. Tan, and G.-P. Li, ``{Shadow images and observed
  luminosity of the Bardeen black hole surrounded by different accretions *},''
  {\em Chin. Phys. C}, vol.~46, no.~8, p.~085106, 2022.

\bibitem{47}
J.~Peng, M.~Guo, and X.-H. Feng, ``{Influence of quantum correction on black
  hole shadows, photon rings, and lensing rings},'' {\em Chin. Phys. C},
  vol.~45, no.~8, p.~085103, 2021.

\bibitem{48}
X.-X. Zeng, G.-P. Li, and K.-J. He, ``{The shadows and observational appearance
  of a noncommutative black hole surrounded by various profiles of
  accretions},'' {\em Nucl. Phys. B}, vol.~974, p.~115639, 2022.

\bibitem{49}
X.-X. Zeng, L.-F. Li, P.~Li, B.~Liang, and P.~Xu, ``{Holographic images of a
  charged black hole in Lorentz symmetry breaking massive gravity},'' {\em Sci.
  China Phys. Mech. Astron.}, vol.~68, no.~2, p.~220412, 2025.

\bibitem{AraujoFilho:2024mvz}
A.~A. Ara{\'u}jo~Filho, N.~Heidari, and A.~{\"O}vg{\"u}n, ``{Geodesics,
  accretion disk, gravitational lensing, time delay, and effects on neutrinos
  induced by a non-commutative black hole},'' {\em JCAP}, vol.~06, p.~062,
  2025.

\bibitem{Lambiase:2023zeo}
G.~Lambiase, L.~Mastrototaro, R.~C. Pantig, and A.~Ovgun, ``{Probing
  Schwarzschild-like black holes in metric-affine bumblebee gravity with
  accretion disk, deflection angle, greybody bounds, and neutrino
  propagation},'' {\em JCAP}, vol.~12, p.~026, 2023.

\bibitem{Macedo:2025ipc}
C.~F.~B. Macedo, J.~L. Rosa, D.~Rubiera-Garcia, and A.~Rueda, ``{Multi-photon
  ring structure of reflection-asymmetric traversable thin-shell wormholes},''
  10 2025.

\bibitem{Olmo:2025ctf}
G.~J. Olmo, J.~L. Rosa, D.~Rubiera-Garcia, A.~Rueda, and
  D.~S{\'a}ez-Chill{\'o}n~G{\'o}mez, ``{Shadows from thin accretion disks of
  parametrized black hole solutions},'' {\em Phys. Rev. D}, vol.~112, no.~8,
  p.~084059, 2025.

\bibitem{Rosa:2024eva}
J.~L. Rosa, J.~Pelle, and D.~P{\'e}rez, ``{Accretion disks and relativistic
  line broadening in boson star spacetimes},'' {\em Phys. Rev. D}, vol.~110,
  no.~8, p.~084068, 2024.

\bibitem{Macedo:2024qky}
C.~F.~B. Macedo, J.~L. Rosa, and D.~Rubiera-Garcia, ``{Optical appearance of
  black holes surrounded by a dark matter halo},'' {\em JCAP}, vol.~07, p.~046,
  2024.

\bibitem{synge1966escape}
J.~Synge, ``The escape of photons from gravitationally intense stars,'' {\em
  Monthly Notices of the Royal Astronomical Society}, vol.~131, no.~3,
  pp.~463--466, 1966.

\bibitem{bardeen1972rotating}
J.~M. Bardeen, W.~H. Press, and S.~A. Teukolsky, ``Rotating black holes:
  locally nonrotating frames, energy extraction, and scalar synchrotron
  radiation,'' {\em Astrophysical Journal, Vol. 178, pp. 347-370 (1972)},
  vol.~178, pp.~347--370, 1972.

\bibitem{gralla2020lensing}
S.~E. Gralla and A.~Lupsasca, ``Lensing by kerr black holes,'' {\em Physical
  Review D}, vol.~101, no.~4, p.~044031, 2020.

\bibitem{shaikh2019shadows}
R.~Shaikh, P.~Kocherlakota, R.~Narayan, and P.~S. Joshi, ``Shadows of
  spherically symmetric black holes and naked singularities,'' {\em Monthly
  Notices of the Royal Astronomical Society}, vol.~482, no.~1, pp.~52--64,
  2019.

\bibitem{he2022shadow}
K.-J. He, S.~Guo, S.-C. Tan, and G.-P. Li, ``Shadow images and observed
  luminosity of the bardeen black hole surrounded by different accretions,''
  {\em Chinese Physics C}, vol.~46, no.~8, p.~085106, 2022.

\bibitem{li2021shadows}
G.-P. Li and K.-J. He, ``Shadows and rings of the kehagias-sfetsos black hole
  surrounded by thin disk accretion,'' {\em Journal of Cosmology and
  Astroparticle Physics}, vol.~2021, no.~06, p.~037, 2021.

\bibitem{shi2024shadow}
Y.~Shi and H.~Cheng, ``The shadow and gamma-ray bursts of a schwarzschild black
  hole in asymptotic safety,'' {\em Communications in Theoretical Physics},
  vol.~77, no.~2, p.~025401, 2024.

\bibitem{feng2025shadow}
X.-H. Feng and G.-Y. Zhang, ``Shadow and quasi-normal modes of
  schwarzschild-hernquist black hole,'' {\em arXiv preprint arXiv:2509.04001},
  2025.

\bibitem{amir2018shadows}
M.~Amir, B.~P. Singh, and S.~G. Ghosh, ``Shadows of rotating five-dimensional
  charged emcs black holes,'' {\em The European Physical Journal C}, vol.~78,
  no.~5, p.~399, 2018.

\bibitem{eiroa2018shadow}
E.~F. Eiroa and C.~M. Sendra, ``Shadow cast by rotating braneworld black holes
  with a cosmological constant,'' {\em The European Physical Journal C},
  vol.~78, no.~2, p.~91, 2018.

\bibitem{li2021observational}
G.-P. Li and K.-J. He, ``Observational appearances of af (r) global monopole
  black hole illuminated by various accretions,'' {\em The European Physical
  Journal C}, vol.~81, no.~11, p.~1018, 2021.

\bibitem{fathi2023observational}
M.~Fathi and N.~Cruz, ``Observational signatures of a static f (r) black hole
  with thin accretion disk,'' {\em The European Physical Journal C}, vol.~83,
  no.~12, p.~1160, 2023.

\bibitem{gralla2019black}
S.~E. Gralla, D.~E. Holz, and R.~M. Wald, ``Black hole shadows, photon rings,
  and lensing rings,'' {\em Physical Review D}, vol.~100, no.~2, p.~024018,
  2019.

\bibitem{zeng2020influence}
X.-X. Zeng and H.-Q. Zhang, ``Influence of quintessence dark energy on the
  shadow of black hole,'' {\em The European Physical Journal C}, vol.~80,
  no.~11, p.~1058, 2020.

\bibitem{zeng2020shadows}
X.-X. Zeng, H.-Q. Zhang, and H.~Zhang, ``Shadows and photon spheres with
  spherical accretions in the four-dimensional gauss--bonnet black hole,'' {\em
  The European Physical Journal C}, vol.~80, no.~9, p.~872, 2020.

\bibitem{zeng2022shadows}
X.-X. Zeng, G.-P. Li, and K.-J. He, ``The shadows and observational appearance
  of a noncommutative black hole surrounded by various profiles of
  accretions,'' {\em Nuclear Physics B}, vol.~974, p.~115639, 2022.

\bibitem{zeng2025holographic}
X.-X. Zeng, L.-F. Li, P.~Li, B.~Liang, and P.~Xu, ``Holographic images of a
  charged black hole in lorentz symmetry breaking massive gravity,'' {\em
  Science China Physics, Mechanics \& Astronomy}, vol.~68, no.~2, p.~220412,
  2025.

\bibitem{meng2023images}
Y.~Meng, X.-M. Kuang, X.-J. Wang, B.~Wang, and J.-P. Wu, ``Images from disk and
  spherical accretions of hairy schwarzschild black holes,'' {\em Physical
  Review D}, vol.~108, no.~6, p.~064013, 2023.

\bibitem{wang2022optical}
H.-M. Wang, Z.-C. Lin, and S.-W. Wei, ``Optical appearance of
  einstein-{\ae}ther black hole surrounded by thin disk,'' {\em Nuclear Physics
  B}, vol.~985, p.~116026, 2022.

\bibitem{yang2023shadow}
J.~Yang, C.~Zhang, and Y.~Ma, ``Shadow and stability of quantum-corrected black
  holes,'' {\em The European Physical Journal C}, vol.~83, no.~7, p.~619, 2023.

\bibitem{gan2021photon}
Q.~Gan, P.~Wang, H.~Wu, and H.~Yang, ``Photon ring and observational appearance
  of a hairy black hole,'' {\em Physical Review D}, vol.~104, no.~4, p.~044049,
  2021.

\bibitem{guo2022gravitational}
G.~Guo, X.~Jiang, P.~Wang, and H.~Wu, ``Gravitational lensing by black holes
  with multiple photon spheres,'' {\em Physical Review D}, vol.~105, no.~12,
  p.~124064, 2022.

\bibitem{chen2022appearance}
Y.~Chen, G.~Guo, P.~Wang, H.~Wu, and H.~Yang, ``Appearance of an infalling star
  in black holes with multiple photon spheres,'' {\em Science China Physics,
  Mechanics \& Astronomy}, vol.~65, no.~12, p.~120412, 2022.

\bibitem{jaroszynski1997optics}
M.~Jaroszynski and A.~Kurpiewski, ``Optics near kerr black holes: spectra of
  advection dominated accretion flows,'' {\em arXiv preprint astro-ph/9705044},
  1997.

\bibitem{bambi2013can}
C.~Bambi, ``Can the supermassive objects at the centers of galaxies be
  traversable wormholes? {The} first test of strong gravity for mm/sub-mm very
  long baseline interferometry facilities,'' {\em Physical Review
  D—Particles, Fields, Gravitation, and Cosmology}, vol.~87, no.~10,
  p.~107501, 2013.

\end{thebibliography}

\end{document}